\documentclass[journal]{IEEEtran}
\PassOptionsToPackage{ruled}{algorithm2e}
\usepackage[T1]{fontenc}
\usepackage[latin9]{inputenc}
\usepackage{color}
\usepackage{algorithm2e}
\usepackage{amsmath}
\usepackage{amsthm}
\usepackage{amssymb}
\usepackage{graphicx}
\usepackage[unicode=true,
 bookmarks=true,bookmarksnumbered=true,bookmarksopen=true,bookmarksopenlevel=1,
 breaklinks=false,pdfborder={0 0 0},pdfborderstyle={},backref=false,colorlinks=false]
 {hyperref}
\hypersetup{pdftitle={Your Title},
 pdfauthor={Your Name},
 pdfpagelayout=OneColumn, pdfnewwindow=true, pdfstartview=XYZ, plainpages=false}

\makeatletter

\providecommand{\tabularnewline}{\\}

\theoremstyle{plain}
\newtheorem{thm}{\protect\theoremname}
\theoremstyle{remark}
\newtheorem{rem}[thm]{\protect\remarkname}

\usepackage[caption=false,font=footnotesize]{subfig}
\DeclareMathOperator{\maximize}{maximize}

\DeclareMathOperator{\st}{subject~to}
\DeclareMathOperator{\diag}{diag}

\DeclareMathOperator{\tr}{Tr}
\DeclareMathOperator{\rank}{rank}
\DeclareMathOperator{\vect}{vec}

\newcommand{\herm}{^{{\dagger}}}

\newcommand{\algref}[1]{\textbf{Algorithm~\ref{#1}}}

\usepackage{cite}
\usepackage{acronym}
\AtBeginDocument{\acrodef{AO}{alternating optimization}
\acrodef{APGM}{alternating projected gradient method}
\acrodef{APM}{accelerated proximal gradient method}
\acrodef{AP}{access point}
\acrodef{ASP}{antenna separation product}
\acrodef{AWGN}{additive white Gaussian noise}
\acrodef{BC}{broadcast channel}
\acrodef{BCM}{block coordinate maximization}
\acrodef{BEP}{bit error probability}
\acrodef{BER}{bit error rate}
\acrodef{BF-MIMO}[BF\mbox{-}MIMO]{beamforming MIMO}
\acrodef{BF}{beamforming}
\acrodef{BS}{base station}
\acrodef{bpcu}{bits per channel use}
\acrodef{CP}{cyclic prefix}
\acrodef{CPU}{central processing unit}
\acrodef{CR}{cutoff rate}
\acrodef{CSI}{channel state information}
\acrodef{CSIR}{channel state information at RX}
\acrodef{SSK}{space shift keying}
\acrodef{CSIT}{channel state information at TX}
\acrodef{DCMC}{discrete\mbox{-}input continuous\mbox{-}output memoryless channel}
\acrodef{DFT}{discrete Fourier transform}
\acrodef{DL-TR-GSM}{dual-layered transmit-receive \acl{GSM}}
\acrodef{DLT}{dual-layered transmission}
\acrodef{DMA}{dynamic metasurface antenna}
\acrodef{DOA}{direction of arrival}
\acrodef{DoF}{degrees of freedom}
\acrodef{DNN}{deep neural networks}
\acrodef{DPC}{dirty paper coding}
\acrodef{DRL}{deep reinforcement learning}
\acrodef{EE}{energy efficiency}
\acrodef{EGC}{equal gain combining}
\acrodef{EM}{electromagnetic}
\acrodef{EVD}{eigenvalue decomposition}
\acrodef{FPGA}{field programmable gate array}
\acrodef{FSPL}{free space path loss}
\acrodef{FFT}{fast Fourier transform}
\acrodef{FDE}{frequency domain equalization}
\acrodef{GRSM}{generalized \acl{RSM}}
\acrodef{GSM}{generalized \acl{SM}}
\acrodef{HMIMO}{holographic MIMO}
\acrodef{IFFT}{invserse fast Fourier transform}
\acrodef{ICI}{inter-channel interference}
\acrodef{iid}[i.i.d.]{independent and identically distributed}
\acrodef{IMT}{International Mobile Telecommunications}
\acrodef{IQ}{in\mbox{-}phase and quadrature}
\acrodef{ISI}{intersymbol interference}
\acrodef{ISI-free}[ISI\mbox{-}free]{intersymbol interference free}
\acrodef{LIS}{large intelligent surface}
\acrodef{LOS}{line\mbox{-}of\mbox{-}sight}
\acrodef{LP}{linear precoding}
\acrodef{KKT}{Karush\mbox{-}Kuhn\mbox{-}Tucker} 
\acrodef{MAC}{multiple-access channel}
\acrodef{mmWave}{millimeter-wave}
\acrodef{MI}{mutual information}
\acrodef{MIMO}{multiple\mbox{-}input multiple\mbox{-}output}
\acrodef{mMIMO}{massive MIMO}
\acrodef{MISO}{multiple\mbox{-}input single\mbox{-}output}
\acrodef{ML}{maximum likelihood}
\acrodef{MRC}{maximal ratio combining}
\acrodef{MMSE}{minimum mean square error}
\acrodef{MU-TR-GSM}{multiuser transmit-receive  \acl{GSM} }
\acrodef{NCSIT}{no channel state information at TX}
\acrodef{NLOS}{non\mbox{-}\acs{LOS}} 
\acrodef{NOMA}{non-orthogonal multiple access}
\acrodef{OFDM}{orthogonal frequency division multiplexing}
\acrodef{OFDMA}{orthogonal frequency division multiple access}
\acrodef{umMIMO}{ultra-massive MIMO}
\acrodef{PA}{power amplifier}
\acrodef{PAE}{power added efficiency}
\acrodef{PAPR}{peak\mbox{-}to\mbox{-}average power ratio}
\acrodef{PDF}{probability density function}
\acrodef{PEP}{pairwise error probability}
\acrodefplural{PEP}{pairwise error probabilities}
\acrodef{PGM}{projected gradient method}
\acrodef{PMP}{probability mass function}
\acrodef{PSM}{precoding-aided spatial modulation}
\acrodef{QSM}{quadrature spatial modulation}
\acrodef{RC}{reorganization computation}
\acrodef{RF}{radio frequency}
\acrodef{RHS}{right-hand side}
\acrodef{RIS}{reconfigurable intelligent surface}
\acrodef{RSM}{receive spatial modulation}
\acrodef{RX}{receiver}
\acrodef{SE}{spectral efficiency}
\acrodef{SEP}{symbol error probability}
\acrodef{SER}{symbol error rate}
\acrodef{SIC}{successive interference cancellation}
\acrodef{SIM}{stacked intelligent metasurface}
\acrodef{SINR}{signal-to-interference-plus-noise ratio}
\acrodef{SISO}{single-input single-output}
\acrodef{SM}{spatial modulation}
\acrodef{SMX-MIMO}[SMX\mbox{-}MIMO]{spatial multiplexing MIMO}
\acrodef{SMX}{spatial multiplexing}
\acrodef{SNR}{signal-to-noise ratio}
\acrodef{SC}{single carrier}
\acrodef{SCA}{successive convex approximation}
\acrodef{SVD}{singular value decomposition}
\acrodef{SPST}{single pole single-throw}
\acrodef{SU}{secondary user}
\acrodef{TDE}{time domain equalization}
\acrodef{TX}{transmitter}
\acrodef{ULA}{uniform linear array}
\acrodef{URA}{uniform rectangular array}
\acrodef{VGA}{variable gain amplifier}
\acrodef{wrt}[w.r.t.]{with respect to}
\acrodef{ZF}{zero-forcing}
\acrodef{ZMCG}{zero-mean complex Gaussian}

}

\ifdefined\showcaptionsetup
 \PassOptionsToPackage{caption=false}{subfig}
\fi
\usepackage{subfig}
\makeatother

\providecommand{\remarkname}{Remark}
\providecommand{\theoremname}{Theorem}

\begin{document}
\title{Energy-Efficient Designs for SIM-Based Broadcast MIMO Systems}
\author{Nemanja~Stefan~Perovi\'c,~\IEEEmembership{Member,~IEEE}, Eduard~E.~Bahingayi,~\IEEEmembership{Member,~IEEE}
and~Le-Nam~Tran,~\IEEEmembership{Senior~Member,~IEEE}\thanks{This work was supported by Taighde \'Eireann - Research Ireland under
Grant 22/US/3847 and Grant 13/RC/2077\_P2. }\thanks{Nemanja~Stefan Perovi\'c was with Universit\'e Paris-Saclay, CNRS,
CentraleSup\'elec, Laboratoire des Signaux et Syst\`emes, 3 Rue
Joliot-Curie, 91192 Gif-sur-Yvette, France. He is now with the Intelligent
Wireless Communication (i-Wic) Laboratory, Institute of Communications
Engineering, National Sun Yat-sen University, Kaohsiung 80424, Taiwan,
R.O.C. (Email: n.s.perovic@mail.nsysu.edu.tw).}\thanks{Eduard E. Bahingayi was with University College Dublin, Belfield,
Dublin, Ireland. He is now with the Engineering Division, New York
University Abu Dhabi (NYUAD), Abu Dhabi, UAE, 129188 (Email: eeb9783@nyu.edu).}\thanks{Le-Nam~Tran is with the School of Electrical and Electronic Engineering,
University College Dublin, Belfield, Dublin 4, D04~V1W8, Ireland
(Email: nam.tran@ucd.ie).}}
\maketitle
\begin{abstract}
Stacked intelligent metasurface (SIM)\acused{SIM}, which consists
of multiple layers of intelligent metasurfaces, is emerging as a promising
solution for future wireless communication systems. In this timely
context, we focus on broadcast \ac{MIMO} systems and aim to characterize
their \ac{EE} performance. To explore the potential of SIM, we consider
both \ac{DPC} and \ac{LP} and formulate the corresponding \ac{EE}
maximization problems. For \ac{DPC}, we employ the \ac{BC}-\ac{MAC}
duality to obtain an equivalent problem, and optimize users' covariance
matrices using the \ac{SCA} and Dinkelbach\textquoteright s methods.
Since the phase shift optimization problem of the SIM meta-elements
is one of extremely large size, we adopt a conventional projected
gradient-based method for its simplicity. A similar approach is followed
for the case of LP. Simulation results show that the proposed optimization
methods for the considered SIM-based systems can significantly improve
the EE, compared to conventional counterparts.  Also, we demonstrate that the number of SIM meta-elements 
and their distribution across the SIM layers have a significant impact on both the achievable sum-rate and EE performance.\acresetall{}
\end{abstract}


\section{Introduction}

\bstctlcite{BSTcontrol}The framework for the future development of
\ac{IMT} for 2030 highlights sustainability as a fundamental goal
for future communication systems, with the aim of minimizing environmental
impact through efficient resource usage, reduced power consumption,
and lower greenhouse gas emissions \cite{recommendation2023framework}.
This has made energy-efficient wireless communications a key research
focus. At the same time, global mobile network data traffic is projected
to reach 563 exabytes (EBs) by 2029 \cite{ericssonMobileData}, which
requires the evolution of network technologies to meet the increasing
demand. For example, conventional \ac{MIMO} systems are advancing
toward \ac{mMIMO} and \ac{umMIMO} systems. However, these technologies
face challenges in \ac{EE} due to the substantial power consumption
required to support a large number of \ac{RF} chains.

A promising solution to address these challenges is the use of \acp{RIS},
which consist of a large number of programmable metamaterial elements
capable of dynamically tuning their \ac{EM} properties. In this way,
RISs can modify the incoming waves in a programmable and controllable
manner \cite{di2020smart}. This capability allows \acp{RIS} to simultaneously
improve multiple performance metrics, such as spectrum efficiency,
\ac{EE}, and coverage. Unfortunately, the significant path-loss of
RIS-assisted links greatly limits the potential \ac{EE} gains from
\ac{RIS} deployment.

To address the limitations of \ac{RIS}, several innovative metamaterial-based
antenna technologies such as holographic radio, \acp{DMA}, and \acp{SIM}
have been proposed. Holographic radio, also known as \ac{HMIMO},
achieves high directive gain, \ac{SE} and \ac{EE} by combining densely
packing sub-wavelength metamaterial antenna elements with holographic
techniques\cite{gong2023holographic,zappone2022energy}. Similarly,
DMAs consist of multiple microstrips, each composed of a multitude
of metamaterial radiating elements and connected to a single RF chain
\cite{shlezinger2021dynamic}. This design offers better EE than hybrid
analog-to-digital (A/D) architectures by eliminating the need for
additional power to support numerous phase shifters \cite{you2022energy}.
However, both \acp{HMIMO} and DMAs, as single-layer structures, often
require a very large number of elements due to hardware constraints
that limit the number of tunable amplitudes/phases.

In contrast, \acp{SIM} represent the latest advancement in metamaterial-based
antenna technologies. Composed of multiple parallel metasurface layers
with nearly passive, programmable meta-elements, \acp{SIM} can be
integrated with conventional transceivers using a small number of
active antennas, offering a low-cost and energy-efficient solution.
Inspired by \ac{DNN}, SIMs perform signal processing tasks, such
as transmit precoding and receive combining, directly in the \ac{EM}
domain when properly controlled and programmed \cite{liu2022programmable}.
The close spacing between active antennas and SIM layers significantly
reduces the path-loss and eliminates the multiplicative path-loss
effects seen in RIS-aided systems. By utilizing whole metasurfaces
to achieve array gain and suppress co-channel interference in the
\ac{EM} wave domain, \acp{SIM} reduce transmission power requirements
while significantly enhancing \ac{EE} and overall system performance,
all with minimal hardware complexity.

Recent studies have investigated various applications of SIMs. In
\cite{an2024two}, SIMs were exploited to implement a 2D \ac{DFT}
for \ac{DOA} estimation. A hybrid channel estimator combining wave
and digital domain processing was proposed in \cite{nadeem2023hybrid}.
However, it requires perfect orthogonality between user pilot sequences,
which is hard to achieve. In \cite{wang2024multi}, the authors minimized
the Cramer-Rao bound (CRB) for target estimation in an ISAC system
with a SIM. Specifically, using an experimental SIM platform, they
showed power variations across layers but without any physical explanation.
A general path-loss model for SIM-assisted systems was developed in
\cite{hassan2024efficient}, based on which, algorithms aimed at maximizing
the received power were derived. This design can generate complex
radiation patterns but requires many layers for discrete phase shifts.
Research on sum-rate maximization includes studies on SIM-assisted
downlink MISO channels \cite{an2023stackedmulti} and multi-user systems
using statistical \ac{CSI} \cite{papazafeiropoulos2024achievableStatistical}.
In particular, it was shown that using statistical \ac{CSI} yields
slightly worse performance while achieving lower computational and
processing overhead. In \cite{liu2024drl}, \ac{DRL} was shown to
outperform conventional optimization approaches for maximizing the
sum rate in SIM-assisted communication systems. In uplink SIM-based
cell-free MIMO systems, rate optimization using distributed signal
processing was studied in \cite{li2024stacked}. In this scenario,
each \ac{AP} locally detects user information, which is then combined
by the \ac{CPU} to recover the final data, though this process can
demand substantial link resources in practice.

The integration of SIMs with transmitters and receivers into a SIM-based
\ac{HMIMO} system, capable of performing signal precoding and combining
in the wave domain, was proposed in \cite{an2023stackedholo}. This
system introduced a channel fitting approach that enables the SIM-based
HMIMO system to achieve significant channel capacity gains compared
to mMIMO and RIS-assisted counterparts. An approach for the mutual
information maximization in a SIM-based HMIMO system with discrete
signaling was presented in \cite{perovic2024mutual}, using the cutoff
rate as an alternative metric. This study demonstrates that incorporating
even a small-scale digital precoder into the system can substantially
increase its mutual information.

Despite extensive research on SIM-assisted systems, the \ac{EE} analysis
of SIM-assisted MIMO systems remains largely unexplored. Notably,
EE optimization inherently includes \ac{SE} maximization and transmit
power minimization as special cases. Motivated by this gap, we aim
to maximize the \ac{EE} in SIM-aided broadcast systems using \ac{DPC}
and \ac{LP}. To solve the formulated EE maximization (EEmax) problems,
we blend multiple optimization techniques to derive efficient problem-driven
algorithms, leveraging the special structure of the considered problems
and suitable optimization tools. For DPC, we formulate a joint optimization
problem of the covariance matrix of the transmitted signal and the
phase shifts of the SIM meta-elements. In the more practical case
of \ac{LP}, we consider a similar joint optimization problem for
transmit precoding and SIM phase shifts. For both scenarios, we assume
a limited total power budget and impose unit modulus constraints on
the SIM meta-elements. Furthermore, perfect CSI is assumed at the
BS to explore the full EE potential of SIM-based systems. Note that
we do not optimize the number of active transmit antennas, which is
left for future work. The main contributions of this paper are summarized
as follows:
\begin{itemize}
\item For DPC, we exploit the well-known Gaussian MIMO \ac{BC}-\ac{MAC}
duality to reformulate the EEmax problem in terms of users\textquoteright{}
covariance matrices in the MAC and the phase shifts of the SIM meta-elements.
Within the adopted \ac{AO} framework, we present an efficient solution
for optimizing the covariance matrices, using a tight concave lower
bound of the achievable sum-rate via the \ac{SCA} method. Dinkelbach\textquoteright s
method is then applied to obtain closed-form expressions for the optimal
covariance matrices. Our complexity analysis demonstrates that our
proposed method has significantly lower complexity compared to an
existing solution \cite{xu2013energy}. For optimizing the SIM phase
shifts, we employ a projected gradient-based method to update all
SIM layers in parallel. This approach is viable considering the large
size of the considered problem. In this context, we derive closed-form
expressions for the complex-valued gradients involved.
\item For LP, we leverage a recent result for the sum-rate maximization
that allows for reformulating the considered EEmax problem as an equivalent
one, with a greatly reduced dimension. Following this important step,
we again invoke the SCA method to derive a quadratic lower bound of
the achievable sum-rate and approximate the EEmax problem as a concave
fractional program. Next, we apply Dinkelbach\textquoteright s method
to solve the resulting problem, obtaining optimal users' precoders
in closed-form expressions. Similar to the DPC-based scheme, the phase
shifts of the SIM meta-elements in this setting are optimized in parallel
using a conventional projected gradient-based method.
\item We present efficient implementations of the proposed algorithms, analyze
their computational complexity in terms of the number of complex multiplications,
and mathematically prove their convergence.
\item We show through simulation results that the proposed algorithms substantially
increase the EE in SIM-aided broadcast communication systems, with
greater improvements observed for DPC. The results highlight the importance
of the proposed precoding schemes in mitigating multi-user interference,
especially in systems with a large number of users. We also provide
several valuable insights into the design and performance of energy-efficient
SIM-based broadcast MIMO systems. First, the EE increases logarithmically
with the number of meta-elements per SIM layer, with a nearly double
gain when increasing from 49 to 100 compared to 100 to 196. Second,
SIM-aided systems with a low number of meta-elements may have lower
EE than conventional MIMO systems without SIM integration. Third,
a few SIM layers suffice to balance SE gains and power consumption
for optimal EE. Fourth, in SIM-aided broadcast systems \emph{without
digital precoding}, optimal EE is achieved by activating a subset
of transmit antennas, each transmitting independent data streams.
Lastly, at least 3 bits per meta-element are required to keep EE degradation
from quantization errors within acceptable limits.
\end{itemize}
The mathematical notation used in this paper is summarized in Table
\ref{tab:Notation}.
\begin{table}[t]
\centering{}\textcolor{black}{\caption{Notation description. \label{tab:Notation}}
}%
\begin{tabular}{|c|c|}
\hline 
\textcolor{black}{Symbol} & \textcolor{black}{Description}\tabularnewline
\hline 
\hline 
\textcolor{black}{$\mathbf{H}$} & \textcolor{black}{Matrix}\tabularnewline
\hline 
\textcolor{black}{$\mathbf{h}$} & \textcolor{black}{Row/column vector}\tabularnewline
\hline 
\textcolor{black}{$\mathbb{C}^{m\times n}$} & \textcolor{black}{Space of $m\times n$ complex matrices}\tabularnewline
\hline 
\textcolor{black}{$\mathbf{H}^{T}$} & \textcolor{black}{Transpose of $\mathbf{H}$}\tabularnewline
\hline 
\textcolor{black}{$\mathbf{H}^{H}$} & \textcolor{black}{Hermitian transpose of $\mathbf{H}$}\tabularnewline
\hline 
\textcolor{black}{$|\mathbf{H}|$} & \textcolor{black}{Determinant of $\mathbf{H}$}\tabularnewline
\hline 
\textcolor{black}{$\tr(\mathbf{H})$} & \textcolor{black}{Trace of $\mathbf{H}$}\tabularnewline
\hline 
\textcolor{black}{$\log_{2}(\cdot)$} & \textcolor{black}{Binary logarithm}\tabularnewline
\hline 
\textcolor{black}{$\ln(\cdot)$} & \textcolor{black}{Natural logarithm}\tabularnewline
\hline 
\textcolor{black}{$(\cdot)^{+}$} & \textcolor{black}{Pseudo-inverse operation}\tabularnewline
\hline 
\textcolor{black}{$\left\lfloor \cdot\right\rfloor $} & \textcolor{black}{Floor operation}\tabularnewline
\hline 
\textcolor{black}{$\left(\cdot\right)^{\ast}$} & \textcolor{black}{Complex conjugate}\tabularnewline
\hline 
\textcolor{black}{$\left\Vert \mathbf{H}\right\Vert $} & \textcolor{black}{Frobenius norm}\tabularnewline
\hline 
\textcolor{black}{$\vect_{d}(\mathbf{H})$} & \textcolor{black}{Vector consisting of the diagonal elements of $\mathbf{H}$}\tabularnewline
\hline 
\textcolor{black}{$\mathbf{A}\succeq(\succ)\mathbf{B}$} & \textcolor{black}{$\mathbf{A}-\mathbf{B}$ is positive semidefinite
(definite)}\tabularnewline
\hline 
\textcolor{black}{$\mathbf{I}$} & \textcolor{black}{Identity matrix}\tabularnewline
\hline 
\textcolor{black}{$\Re(\mathbf{x})$} & \textcolor{black}{Real part of $\mathbf{x}$}\tabularnewline
\hline 
\textcolor{black}{$\Im(\mathbf{x})$} & \textcolor{black}{Imaginary part of $\mathbf{x}$}\tabularnewline
\hline 
\textcolor{black}{$\diag(\mathbf{x})$} & \textcolor{black}{Matrix with the elements of $\mathbf{x}$ on the
main diagonal}\tabularnewline
\hline 
\textcolor{black}{$\mathcal{CN}(\cdot,\cdot$)} & \textcolor{black}{Circularly symmetric complex Gaussian distribution}\tabularnewline
\hline 
\textcolor{black}{$|x|$} & \textcolor{black}{Modulus of the number $x$}\tabularnewline
\hline 
\textcolor{black}{|$\mathbf{H}$|} & \textcolor{black}{Determinant of $\mathbf{H}$}\tabularnewline
\hline 
\textcolor{black}{$\nabla_{\mathbf{x}}f(\cdot)$} & \textcolor{black}{Complex gradient of $f(\cdot)$ \ac{wrt} $\mathbf{x}^{\ast}$}\tabularnewline
\hline 
\end{tabular}
\end{table}

\section{System Model and Problem Formulation}

\subsection{System Model}

\begin{figure}[t]
\begin{centering}
\subfloat[\textcolor{black}{Conventional broadcast system.}]{\centering{}\includegraphics[scale=0.5]{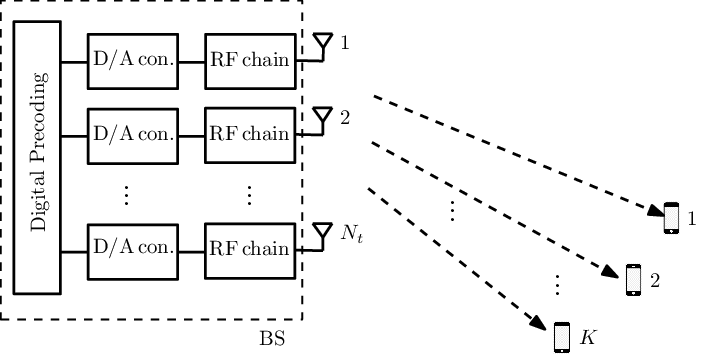}}
\par\end{centering}
\centering{}\subfloat[\textcolor{black}{SIM-aided broadcast system.}]{\centering{}\includegraphics[scale=0.5]{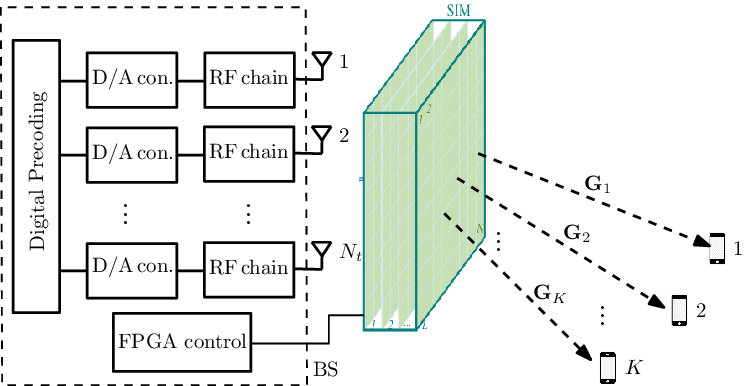}}\caption{\textcolor{black}{System level diagrams of conventional and SIM-aided
broadcast systems.\label{fig:System-level-diagrams}}}
\end{figure}

\textcolor{black}{System level diagrams of conventional and SIM-aided
broadcast systems are shown in Fig. \ref{fig:System-level-diagrams}.
In both cases, }we consider a multi-user broadcast system in which
a BS with $N_{t}$ transmit antennas communicates with $K$ users,
each equipped $N_{r}$ receive antennas\textcolor{blue}{. }\textcolor{black}{The
main difference is in the presence of a SIM at the BS,} which comprises
of $L$ metasurface layers, each containing $N$ meta-elements. The
SIM is controlled by external field programmable gate array (FPGA)
devices, which adjust the phase shifts of individual meta-elements
for wave domain beamforming.\footnote{By considering both SIM (i.e., wave-based precoding) and digital precoding,
our system model is general enough to include the wave-based only
precoding and digital only precoding as special cases.} \textcolor{black}{For the sake of simplicity, we assume that this
device is integrated into the BS hardware and communicates with the
SIM via a dedicated interface.}

The phase shifts of the meta-elements in the $l$-th SIM layer are
represented by the diagonal matrix $\boldsymbol{\mathbf{\Phi}}^{l}=\text{diag}(\boldsymbol{\phi}^{l})=\text{diag}([\phi_{1}^{l}\:\phi_{2}^{l}\:\cdots\:\phi_{N}^{l}]^{T})$,
where $\phi_{n}^{l}=\exp(j\theta_{n}^{l})$ and $\theta_{n}^{l}$
is the phase shift of the $n$-th element in the $l$-th layer. Signal
propagation between two consecutive layers, $l-1$ and $l$, is modeled
by the matrix $\mathbf{W}^{l}\in\mathbb{C}^{N\times N}$ for $l=2,3,\dots,L$.
Specifically, the signal propagation between the $n$-th meta-element
of the $(l-1)$-th and the $m$-th meta-element of $l$-th layer of
the SIM is presented by\cite[Eq. (1)]{lin2018all}
\begin{equation}
[\mathbf{W}^{l}]_{m,n}=\frac{A_{k}\cos\chi_{m,n}}{d_{m,n}}\Bigl(\frac{1}{2\pi d_{m,n}}-\frac{j}{\lambda}\Bigr)e^{j\frac{2\pi d_{m,n}}{\lambda}}\label{eq:RS_equ}
\end{equation}
where $A_{k}$ is the area of each meta-element, $d_{m,n}$ is the
distance between the meta-elements of these two layers, $\chi_{m,n}$
is the angle between the propagation direction and the normal to the
$(l-1)$-th layer, and $\lambda$ is the wavelength. The signal propagation
between the transmit antenna array and the first layer of the SIM
is represented by the matrix $\mathbf{W}^{1}\in\mathbb{C}^{N\times N_{t}}$,
whose elements can be calculated as in (\ref{eq:RS_equ}). 

The end-to-end channel matrix between the BS and the $k$-th user
is expressed as
\begin{equation}
\mathbf{H}_{k}=\ensuremath{\mathbf{G}}_{k}\ensuremath{\mathbf{B}}\in\mathbb{C}^{N_{r}\times N_{t}}
\end{equation}
where $\text{\ensuremath{\mathbf{G}_{k}} \ensuremath{\in\mathbb{C}^{N_{r}\times N}}}$
denotes the channel matrix between the final layer of the SIM and
user $k$, and $\mathbf{B}=\boldsymbol{\mathbf{\Phi}}^{L}\mathbf{W}^{L}\cdots\boldsymbol{\mathbf{\Phi}}^{2}\mathbf{W}^{2}\boldsymbol{\mathbf{\Phi}}^{1}\mathbf{W}^{1}\in\mathbb{C}^{N\times N_{t}}$
is the EM response of the transmit SIM. We assume that $\ensuremath{\mathbf{G}}_{k}$
is perfectly known to the BS to investigate its full theoretical potential.

\subsection{Dirty Paper Coding (DPC)}

To evaluate an upper bound on achievable EE, we first consider DPC,
which achieves the capacity region of a Gaussian MIMO BC. The received
signal at user $k$ is given by
\begin{equation}
\mathbf{y}_{k}=\mathbf{H}_{k}\mathbf{s}_{k}+\sum\nolimits_{j<k}\mathbf{H}_{k}\mathbf{s}_{j}+\sum\nolimits_{j>k}\mathbf{H}_{k}\mathbf{s}_{j}+\mathbf{n}_{k}
\end{equation}
where $\mathbf{H}_{k}\in\mathbb{C}^{N_{r}\times N_{t}}$ is the channel
matrix for user $k$, $\mathbf{s}_{k}\in\mathbb{C}^{N_{t}\times1}$
is the transmitted signal for user $k$. The noise vector $\mathbf{n}_{k}\in\mathbb{C}^{N_{r}\times1}$
consists of \ac{iid} elements that are distributed according to $\mathcal{CN}(0,\sigma^{2})$,
where $\sigma^{2}$ is the noise variance.

The BS designs $\mathbf{s}_{k}$ with full (non-causal) knowledge
of the interference term $\sum_{j<k}\mathbf{H}_{k}\mathbf{s}_{j}$,
which results from transmissions to users $1,2,\dots,k-1$. Therefore,
DPC can eliminate this interference from previously encoded users,
and thus, only the interference caused by users $k+1,\dots,K$ remains.
Based on this, the achievable rate of user $k$ is given by
\begin{equation}
R_{\text{BC},k}(\mathbf{Q},\boldsymbol{\phi})=\ln\frac{\Bigl|\mathbf{I}+\mathbf{H}_{k}\bigl(\sum_{j\geq k}\mathbf{Q}_{j}\bigr)\mathbf{H}_{k}^{H}\Bigr|}{\Bigl|\mathbf{I}+\mathbf{H}_{k}\bigl(\sum_{j>k}\mathbf{Q}_{j}\bigr)\mathbf{H}_{k}^{H}\Bigr|},\label{eq:Rate_BC}
\end{equation}
where, by slight abuse of notation, $\mathbf{H}_{k}$ stands for $\mathbf{H}_{k}/\sigma$
(i.e., $\mathbf{H}_{k}$ is normalized by the square root of the noise
power), $\mathbf{Q}_{k}=\mathbb{E}\bigl\{\mathbf{s}_{k}\mathbf{s}_{k}^{H}\bigr\}\succeq\mathbf{0}$
is the input covariance matrix of user $k$, $\mathbf{Q}=[\mathbf{Q}_{1},\mathbf{Q}_{2},\dots,\mathbf{Q}_{K}]$
and $\boldsymbol{\phi}=[(\boldsymbol{\phi}^{1})^{T},(\boldsymbol{\phi}^{2})^{T},\dots,(\boldsymbol{\phi}^{L})^{T}]^{T}$.
These covariance matrices are constrained by the total power budget
as
\begin{equation}
\sum\nolimits_{k=1}^{K}\tr(\mathbf{Q}_{k})\le P_{\max}\label{eq:pow_const_DPC}
\end{equation}
where $P_{\max}$ is the available transmit power budget.

\subsection{Multi-user MIMO with LP}

Although DPC is a capacity achieving scheme, it has high complexity
due to its nonlinear coding nature. On the other hand, LP is much
simpler to implement in practice. For LP, the transmitted signal is
expressed as
\begin{equation}
\mathbf{x}=\sum\nolimits_{k=1}^{K}\mathbf{P}_{k}\mathbf{s}_{k}
\end{equation}
where $\mathbf{s}_{k}$ is the signal intended for user $k$ and $\mathbf{P}_{k}\in\mathbb{C}^{N_{t}\times N_{r}}$
is the corresponding linear precoder. Thus, the received signal at
user $k$ is given by
\begin{align}
\mathbf{y}_{k}=\mathbf{H}_{k}\mathbf{x}+\mathbf{n}_{k}=\mathbf{H}_{k}\mathbf{P}_{k}\mathbf{s}_{k}+\sum_{j=1,j\neq k}^{K}\mathbf{H}_{k}\mathbf{P}_{j}\mathbf{s}_{j}+\mathbf{n}_{k}.\label{eq:yk}
\end{align}
 By treating the multiuser interference as Gausian noise, the achievable
rate of user $k$ is given by
\begin{align}
\!\!\!R_{\text{L},k}(\mathbf{P},\boldsymbol{\phi})= & \ln\bigg|\mathbf{I}+\mathbf{H}_{k}\mathbf{P}_{k}\mathbf{P}_{k}^{H}\mathbf{H}_{k}^{H}\nonumber \\
 & \times\left(\mathbf{I}+\sum\nolimits_{j=1,j\neq k}^{K}\mathbf{H}_{k}\mathbf{P}_{j}\mathbf{P}_{j}^{H}\mathbf{H}_{k}^{H}\right)^{-1}\bigg|\label{eq:Rate_user}
\end{align}
where $\mathbf{H}_{k}=\mathbf{H}_{k}/\sigma$ and the precoding matrices
$\mathbf{P}=[\mathbf{P}_{1},\mathbf{P}_{2},\dots,\mathbf{P}_{K}]$
have to satisfy the total power constraint:
\begin{equation}
\sum\nolimits_{k=1}^{K}\tr(\mathbf{P}_{k}\mathbf{P}_{k}^{H})\le P_{\max}.\label{eq:pow_constr-LP}
\end{equation}

Note that the achievable rate of user $k$ can alternatively be expressed
using covariance matrices by defining $\mathbf{Q}_{k}^{\prime}=\mathbf{P}_{k}\mathbf{P}_{k}^{H}$.
However, $\mathbf{Q}_{k}^{\prime}$ differs from $\mathbf{Q}_{k}$
for DPC because of differences in the rate expressions. Some existing
solutions impose $\rank(\mathbf{Q}_{k})=N_{r}$ to derive $\mathbf{P}_{k}$,
but we do not adopt this approach because optimizing $\{\mathbf{Q}_{k}^{\prime}\}$
introduces additional computational complexity compared to directly
optimizing~$\{\mathbf{P}_{k}\}$.

\subsection{Problem Formulation}

The objective is to maximize the \ac{EE} of the considered communication
system, which is defined as the ratio of the sum-rate and the total
power consumption. The total power consumption is modeled as $P_{\mathrm{tot}}=P_{t}+N_{t}P_{c}+P_{0}+LNP_{s},$
where
\begin{equation}
P_{t}=\begin{cases}
\sum\nolimits_{k=1}^{K}\tr(\mathbf{Q}_{k}) & \text{DPC}\\
\sum\nolimits_{k=1}^{K}\tr(\mathbf{P}_{k}\mathbf{P}_{k}^{H}) & \text{LP}
\end{cases}
\end{equation}
is the data-dependent transmit signal power, $P_{c}$ is the circuit
power per RF chain (including the appropriate D\textbackslash A converter),
$P_{0}$ is the standby power consumed at the BS, and $P_{s}$ is
the power consumption of the switching circuits (e.g., PIN diodes,
varactor diodes) of each SIM meta-element.\footnote{In this model, we assume that the FPGA control and driving circuits
of the SIM are integrated into the BS hardware, and the power consumption
of these circuits is already included in the BS power consumption,
$P_{0}$.} Since BSs dominate power consumption in mobile networks, user power
consumption is excluded from the EE optimization in this study.

For the DPC-based scheme, the EE maximization (EEmax) problem is stated
as 
\begin{subequations}
\label{eq:opt_problem_BC}
\begin{align}
\underset{\mathbf{Q},\boldsymbol{\phi}}{\max} & \ \eta_{\mathrm{dpc}}=\frac{W\sum_{k=1}^{K}R_{\text{BC},k}(\mathbf{Q},\boldsymbol{\phi})}{\sum_{k=1}^{K}\tr(\mathbf{Q}_{k})+N_{t}P_{c}+P_{0}+LNP_{s}}\label{eq:Obj_BC}\\
\mathrm{s.t.} & \ \left|\phi_{n}^{l}\right|=1,\forall l,n\label{eq:phase constr:DPC}\\
 & \ \sum\nolimits_{k=1}^{K}\tr(\mathbf{Q}_{k})\le P_{\text{max}};\mathbf{Q}_{k}\succeq\mathbf{0},\forall k,\label{eq:pow_constr}
\end{align}
\end{subequations}
where $W$ is the system bandwidth. Similarly, the EEmax problem with
LP is written as 
\begin{subequations}
\label{eq:opt_problem-LP}
\begin{align}
\underset{\mathbf{P},\boldsymbol{\phi}}{\max} & \ \eta_{\mathrm{lp}}=\frac{W\sum_{k=1}^{K}R_{\text{L},k}(\mathbf{P},\boldsymbol{\phi})}{\sum_{k=1}^{K}\tr(\mathbf{P}_{k}\mathbf{P}_{k}^{H})+N_{t}P_{c}+P_{0}+LNP_{s}}\label{eq:Obj_LP}\\
\mathrm{s.t.} & \ \left|\phi_{n}^{l}\right|=1,\forall l,n\label{eq:phase constr:LP}\\
 & \ \sum\nolimits_{k=1}^{K}\tr(\mathbf{P}_{k}\mathbf{P}_{k}^{H})\le P_{\text{max}}.\label{eq:pow_constr_LP}
\end{align}
\end{subequations}
 Note that the equality constraints in (\ref{eq:phase constr:DPC})
and (\ref{eq:phase constr:LP}) are treated element-wise. Since $W$
is constant, it is dropped when solving (\ref{eq:opt_problem_BC})
and (\ref{eq:opt_problem-LP}), but included in simulation results
in Section \ref{sec:Sim-Res}. Next, we present our proposed methods
for solving the above two EEmax problems.

\section{Proposed Solution to DPC-based SIM}

To solve (\ref{eq:opt_problem_BC}), we present an iterative optimization
algorithm that alternates between optimizing the covariance matrices
and the SIM phase shifts, a prevailing method in SIM-related studies.
In particular, for fixed phase shifts, we propose a novel method to
optimize the covariance matrices in parallel, using a closed-form
water-filling solution, which allocates more transmit power to ``stronger''
user channels. This nice result is achieved by applying Dinkelbach\textquoteright s
method to iteratively maximize a quadratic lower bound of the objective
function. The phase shifts of the SIM meta-elements are optimized
by a gradient-based optimization method, which is a viable choice,
considering the extremely large size of the SIM.

\subsection{Covariance Matrix Optimization}

For ease of readability, we briefly outlined the steps for optimizing
covariance matrices as follows:
\begin{itemize}
\item Reformulate (\ref{eq:opt_problem_BC}) using the \acp{BC}-\acp{MAC}
duality to obtain a more tractable equivalent formulation.
\item Derive a quadratic lower bound for the achievable sum-rate under the
SCA framework.
\item Apply Dinkelbach\textquoteright s method to solve the fractional EE
objective, leading to a water-filling solution for the covariance
matrices.
\end{itemize}

\subsubsection{Problem Reformulation}

The EEmax problem in (\ref{eq:Obj_BC}) is challenging to solve since
its objective function is neither convex nor concave with respect
to the optimization variables. To deal with this, we exploit the well-known
duality between \acp{BC} and \acp{MAC} \cite{vishwanath2003duality}\textcolor{blue}{,
}\textcolor{black}{which states that the capacity region of the MIMO
BC is equal to the capacity achieving region of the dual MIMO MAC.
For single-user MIMO systems, this duality means that the capacity
of a downlink channel $\mathbf{H}$ equals that of a dual uplink channel
$\mathbf{H}^{H}$ under the same power constraint, which is easily
understood. Extending this idea to multi-user MIMO systems, for a
given channel between the BS and user $k$, the BC channel $\mathbf{H}_{k}$
corresponds to the dual MAC channel $\mathbf{H}_{k}^{H}$. Similarly,
each transmit covariance matrix $\mathbf{Q}_{k}$ in the BC has as
its counterpart a corresponding dual covariance matrix $\mathbf{S}_{k}$
in the dual MAC.}\footnote{\textcolor{black}{It should be noted that $\mathbf{Q}_{k}$ can not
be directly mapped to $\mathbf{S}_{k}$. Instead, it is required to
compute first $\mathbf{S}_{K},\mathbf{S}_{K-1},\dots,\mathbf{S}_{k+1}$
before obtaining $\mathbf{S}_{k}$ \cite{vishwanath2003duality}.}}\textcolor{black}{{} Accordingly, (\ref{eq:opt_problem_BC}) is equivalent
to the EEmax problem in the dual MAC, which is expressed as}
\begin{subequations}
\textcolor{black}{\label{eq:opt_problem_MAC}
\begin{align}
\!\text{\!\!\ensuremath{\underset{\mathbf{S}}{\maximize}}	} & \quad\frac{\ln\left|\mathbf{I}+\sum_{k=1}^{K}\mathbf{H}_{k}^{H}\mathbf{S}_{k}\mathbf{H}_{k}\right|}{\sum_{k=1}^{K}\tr(\mathbf{S}_{k})+N_{t}P_{c}+P_{0}+LNP_{s}}\label{eq:Obj_MAC}\\
 & \quad\sum\nolimits_{k=1}^{K}\tr(\mathbf{S}_{k})\le P_{\text{max}};\mathbf{S}_{k}\succeq\mathbf{0},\forall k,\label{eq:pow_constr_MAC}
\end{align}
}
\end{subequations}
\textcolor{black}{where we have omitted (\ref{eq:phase constr:DPC})
as the phase shifts are fixed, and the power constraint (\ref{eq:pow_constr_MAC})
has the same maximum power as (\ref{eq:pow_constr})}\textcolor{blue}{.
}To solve (\ref{eq:opt_problem_MAC}) efficiently, we \emph{temporarily}
drop the power constraint (\ref{eq:pow_constr}), which results in
\begin{align}
\!\!\!\text{\ensuremath{\underset{\mathbf{S}_{k}\succeq0}{\maximize}}	}g(\mathbf{S})= & \frac{\ln\left|\mathbf{I}+\sum_{k=1}^{K}\mathbf{H}_{k}^{H}\mathbf{S}_{k}\mathbf{H}_{k}\right|}{\sum_{k=1}^{K}\tr(\mathbf{S}_{k})+N_{t}P_{c}+P_{0}+LNP_{s}}.\label{eq:Opt_unconst}
\end{align}
To appreciate the novelty of our proposed method, we briefly describe
the block-coordinate method proposed in \cite{xu2013energy}, which
optimizes each $\mathbf{S}_{k}$ sequentially, while other variables
are fixed. More precisely, let $\mathbf{S}^{(n)}=[\mathbf{S}_{1}^{(n)},\mathbf{S}_{2}^{(n)},\dots,\mathbf{S}_{K}^{(n)}]$
denote the current iterate. Then the next iterate $\mathbf{S}_{k}^{(n+1)}$
is obtained as
\begin{align}
\mathbf{S}_{k}^{(n+1)} & =\arg\underset{\mathbf{S}_{k}\succeq\mathbf{0}}{\max}\:\:g\bigl(\mathbf{S}_{1}^{(n+1)},\ldots,\mathbf{S}_{k-1}^{(n+1)},\mathbf{S}_{k},\mathbf{S}_{k+1}^{(n)},\ldots,\mathbf{S}_{K}^{(n)}\bigr)\nonumber \\
 & =\arg\underset{\mathbf{S}_{k}\succeq\mathbf{0}}{\max}\:\:\frac{\ln\left|\mathbf{Z}_{k}\right|+\ln\left|\mathbf{I}+\mathbf{T}_{k}^{H}\mathbf{S}_{k}\mathbf{T}_{k}\right|}{p_{k}+\tr(\mathbf{S}_{k})}\label{eq:opt_Sk}
\end{align}
where $\mathbf{Z}_{k}=\mathbf{I}+\sum_{j=1,j\neq k}^{K}\mathbf{H}_{j}^{H}\mathbf{S}_{j}\mathbf{H}_{j}$,
$\mathbf{T}_{k}=\mathbf{H}_{k}\mathbf{Z}_{k}^{-1/2}$, and $p_{k}=\sum_{j=1,j\neq k}^{K}\tr(\mathbf{\mathbf{S}}_{j})+N_{t}P_{c}+P_{0}+LNP_{s}$.
Next, to solve (\ref{eq:opt_Sk}), the authors of \cite{xu2013energy}
applied Dinkelbach\textquoteright s method, which leads to the following
problem 
\begin{equation}
\underset{\mathbf{S}_{k}\succeq\mathbf{0}}{\max}F_{\lambda}\bigl(\mathbf{S}\bigr)\triangleq\ln\left|\mathbf{Z}_{k}\right|+\ln\left|\mathbf{I}+\mathbf{T}_{k}^{H}\mathbf{S}_{k}\mathbf{T}_{k}\right|-\lambda_{D}(p_{k}+\tr(\mathbf{S}_{k}))\label{eq:Dinkl_trans}
\end{equation}
where $\lambda_{D}$ is a non-negative parameter. For a given $\lambda_{D}$,
the above problem can be solved in closed-form \cite{xu2013energy}.
However, such a method requires finding the inverse of $\mathbf{Z}_{k}\in\mathcal{C}^{N_{t}\times N_{t}}$
which has a complexity of $\mathcal{O}(N_{t}^{3})$ in general, and
computing the \ac{SVD} of $\mathbf{T}_{k}$ which has a complexity
of $\mathcal{O}(N_{t}^{2}N_{r})$. Thus, the overall complexity of
the method presented in \cite{xu2013energy} is very high.

\subsubsection{SCA Method: A Tight Lower Bound of Sum Rate}

We now propose a more efficient method by leveraging the fact that
a stationary solution to (\ref{eq:Opt_unconst}) is also globally
optimal since (\ref{eq:Opt_unconst}) is \emph{a concave-convex program}.
This observation motivates us to adopt the SCA framework, which is
typically used to find a stationary solution for nonconvex programs.
To proceed, since $\mathbf{S}_{k}\succeq\mathbf{0}$, we can write
$\mathbf{S}_{k}=\mathbf{U}_{k}^{H}\mathbf{U}_{k}$ where $\mathbf{U}_{k}\in\mathbb{C}^{N_{r}\times N_{r}}$.
Thus, (\ref{eq:Opt_unconst}) is equivalent to the following unconstrained
optimization problem:
\begin{equation}
\text{\ensuremath{\underset{\mathbf{U}_{k}}{\maximize}} }g(\mathbf{U})=\frac{h(\mathbf{U)}}{\sum_{k=1}^{K}\tr(\mathbf{U}_{k}^{H}\mathbf{U}_{k})+N_{t}P_{c}+P_{0}+LNP_{s}}.\label{eq:Opt_unconst-precoder}
\end{equation}
where $h(\mathbf{U)}=\ln\left|\mathbf{I}+\sum_{k=1}^{K}\mathbf{H}_{k}^{H}\mathbf{U}_{k}^{H}\mathbf{U}_{k}\mathbf{H}_{k}\right|$
and $\mathbf{U}=[\mathbf{U}_{1},\mathbf{U}_{2},\dots,\mathbf{U}_{K}]$.
It is easy to see that $h(\mathbf{U)}$ can be equivalently rewritten
as
\begin{subequations}
\begin{gather*}
h(\mathbf{U)}=\sum_{j=1}^{K}\ln\Bigl|\mathbf{I}+\Bigl(\mathbf{I}+\sum_{k=j+1}^{K}\mathbf{H}_{k}^{H}\mathbf{U}_{k}^{H}\mathbf{U}_{k}\mathbf{H}_{k}\Bigr)^{-1}\mathbf{H}_{k}^{H}\mathbf{U}_{k}^{H}\mathbf{U}_{k}\mathbf{H}_{k}\Bigr|\\
=\sum_{j=1}^{K}\ln\Bigl|\mathbf{I}+\mathbf{U}_{j}\mathbf{H}_{j}\Bigl(\mathbf{I}+\sum_{k=j+1}^{K}\mathbf{H}_{k}^{H}\mathbf{U}_{k}^{H}\mathbf{U}_{k}\mathbf{H}_{k}\Bigr)^{-1}\mathbf{H}_{j}^{H}\mathbf{U}_{j}^{H}\Bigr|.
\end{gather*}
\end{subequations}
In fact, the $j$-th term in the sum above is the capacity of user
$j$ in the dual MAC, using successive interference cancellation \cite{vishwanath2003duality}.
As shown shortly, the above reformulation of $h(\mathbf{U)}$ allows
for approximating $h(\mathbf{U)}$ by a ``proper bound'' to obtain
a subproblem that admits a closed-form solution. In this regard, we
recall the following inequality \cite{tam2016successive}:
\begin{gather}
\ln\left|\mathbf{I}+\mathbf{V}\mathbf{Y}^{-1}\mathbf{V}^{H}\right|\ge\ln\left|\mathbf{I}+\hat{\mathbf{V}}\hat{\mathbf{Y}}^{-1}\hat{\mathbf{V}}^{H}\right|-\tr(\hat{\mathbf{V}}\hat{\mathbf{Y}}^{-1}\hat{\mathbf{V}}^{H})\nonumber \\
+2\mathfrak{R}(\tr(\hat{\mathbf{V}}\hat{\mathbf{Y}}^{-1}\mathbf{V}^{H}))-\tr(\mathbf{A}^{H}(\mathbf{V}^{H}\mathbf{V}+\mathbf{Y})),\label{eq:quadractic_approx}
\end{gather}
where $\mathbf{A}=\hat{\mathbf{Y}}^{-1}-(\hat{\mathbf{V}}^{H}\hat{\mathbf{V}}+\hat{\mathbf{Y}})^{-1}$.
We remark that the above inequality holds for arbitrary $\mathbf{V}$,
$\hat{\mathbf{V}}$, $\mathbf{Y}\succ\mathbf{0}$ and $\hat{\mathbf{Y}}\succ\mathbf{0}$,
provided that their sizes are compatible and the equality occurs when
$\mathbf{V}=\hat{\mathbf{V}}$ and $\mathbf{Y}=\hat{\mathbf{Y}}$.
In light of the SCA framework, we denote by $\mathbf{U}_{j}^{(n)}$
the value of $\mathbf{U}_{j}$ after $n$ iterations. Let 
\begin{algorithm}[t]
\caption{Optimization of the covariance matrices in the dual MAC. \label{alg:DPC_cov_mat_opt}}

\SetAlgoNoLine
\DontPrintSemicolon
\LinesNumbered 

\KwIn{$\mathbf{H}$, $\mathbf{S}$, $\lambda^{(0)}\ge0$, $n\leftarrow0$}

\Repeat{ $\lambda^{(n)}-\lambda^{(n-1)}>\epsilon_{D}$}{

\For{$j=1,2,\cdots,K$}{

Calculate $\hat{\mathbf{V}}_{j}^{(n)}$, $\hat{\mathbf{Y}}_{j}^{(n)}$,
$\mathbf{A}_{j}^{(n)}$, $\mathbf{B}_{j}^{(n)}$\;

Decompose $\mathbf{H}_{j}\sum_{l=1}^{j}\mathbf{A}_{l}^{(n)}\mathbf{H}_{j}^{H}=\mathbf{P}_{j}\boldsymbol{\Sigma}_{j}\mathbf{P}_{j}^{H}$\;

}

Solve for $\lambda_{D}$ in (\ref{eq:f_lambdaD}) \;

Compute $\mathbf{U}_{j}^{(n+1)}$ using (\ref{eq:Uj_opt}) for $j=1,2,\cdots,K$
\;

$\lambda^{(n+1)}\leftarrow\lambda_{D}$

$n\leftarrow n+1$\;

}

Calculate all $\mathbf{S}_{k}^{*}=\mathbf{U}_{k}^{(n)H}\mathbf{U}_{k}^{(n)}$
\;

\eIf{ $\sum\nolimits_{k=1}^{K}\tr(\mathbf{S}_{k}^{*})\le P$ }{

$\mathbf{S}_{\text{opt}}=\mathbf{S}^{*}$\;}{

$\mathbf{S}_{\text{opt}}=\hat{\mathbf{S}}$ obtained from \cite{perovic2022maximum}\;

}
\end{algorithm}
\[
\mathbf{V}=\mathbf{U}_{j}\mathbf{H}_{j}\triangleq\mathbf{V}_{j}\in\mathbb{C}^{N_{r}\times N_{t}},
\]
\[
\hat{\mathbf{V}}=\mathbf{U}_{j}^{(n)}\mathbf{H}_{j}\triangleq\hat{\mathbf{V}}_{j}^{(n)}\in\mathbb{C}^{N_{r}\times N_{t}},
\]
\[
\mathbf{Y}=\mathbf{I}+\sum_{k=j+1}^{K}\mathbf{H}_{k}^{H}\mathbf{U}_{k}^{H}\mathbf{U}_{k}\mathbf{H}_{k}=\mathbf{I}+\sum_{k=j+1}^{K}\mathbf{V}_{j}^{H}\mathbf{V}_{j}\triangleq\mathbf{Y}_{j}
\]
 and 
\begin{align*}
\hat{\mathbf{Y}} & =\mathbf{I}+\sum\nolimits_{k=j+1}^{K}\mathbf{H}_{k}^{H}\mathbf{U}_{k}^{(n)H}\mathbf{U}_{k}^{(n)}\mathbf{H}_{k}\\
 & =\mathbf{I}+\sum\nolimits_{k=j+1}^{K}\hat{\mathbf{V}}_{j}^{(n)H}\hat{\mathbf{V}}_{j}^{(n)}\triangleq\hat{\mathbf{Y}}_{j}^{(n)}\in\mathbb{C}^{N_{t}\times N_{t}}.
\end{align*}
Then (\ref{eq:quadractic_approx}) implies
\begin{align}
h(\mathbf{U}) & \geq\bar{h}(\mathbf{U};\mathbf{U}^{(n)})=c^{(n)}+\sum\nolimits_{j=1}^{K}2\mathfrak{R}\Bigl(\tr(\mathbf{B}_{j}^{(n)}\mathbf{U}_{j}^{H})\Bigr)\nonumber \\
 & \quad-\sum\nolimits_{j=1}^{K}\tr(\mathbf{A}_{j}^{(n)H}\sum\nolimits_{k=j}^{K}\mathbf{H}_{k}^{H}\mathbf{U}_{k}^{H}\mathbf{U}_{k}\mathbf{H}_{k}).\label{eq:LB:inequality}
\end{align}
where
\begin{gather*}
c^{(n)}=\sum\nolimits_{j=1}^{K}\Bigg[\ln\Bigl|\mathbf{I}+\hat{\mathbf{V}}_{j}^{(n)}\Bigl(\hat{\mathbf{Y}}_{j}^{(n)}\Bigr)^{-1}\hat{\mathbf{V}}_{j}^{(n)H}\Bigr|\\
-\tr(\hat{\mathbf{V}}_{j}^{(n)}\Bigl(\hat{\mathbf{Y}}_{j}^{(n)}\Bigr)^{-1}\hat{\mathbf{V}}_{j}^{(n)H})-\tr(\mathbf{A}_{j}^{(n)H})\Bigg]\\
=h(\mathbf{U}^{(n)})-\sum\nolimits_{j=1}^{K}\tr(\hat{\mathbf{V}}_{j}^{(n)}\Bigl(\hat{\mathbf{Y}}^{(n)}\Bigr)^{-1}\hat{\mathbf{V}}_{j}^{(n)H})-\tr(\mathbf{A}_{j}^{(n)H})\\
\mathbf{B}_{j}^{(n)}=\bar{\mathbf{B}}_{j}^{(n)}\mathbf{H}_{j}^{H}\in\mathbb{C}^{N_{r}\times N_{r}};\bar{\mathbf{B}}_{j}^{(n)}=\hat{\mathbf{V}}_{j}^{(n)}\Bigl(\hat{\mathbf{Y}}_{j}^{(n)}\Bigr)^{-1}\in\mathbb{C}^{N_{r}\times N_{t}}\\
\mathbf{A}_{j}^{(n)}=\Bigl(\hat{\mathbf{Y}}_{j}^{(n)}\Bigr)^{-1}-\Bigl(\hat{\mathbf{V}}_{j}^{(n)H}\hat{\mathbf{V}}_{j}^{(n)}+\hat{\mathbf{Y}}_{j}^{(n)}\Bigr)^{-1}\\
=\Bigl(\hat{\mathbf{Y}}_{j}^{(n)}\Bigr)^{-1}-\Bigl(\hat{\mathbf{Y}}_{j-1}^{(n)}\Bigr)^{-1}\in\mathbb{C}^{N_{t}\times N_{t}}.
\end{gather*}
Regarding the complexity of the above approximation, the following
remark is in order.
\begin{rem}
Since $\hat{\mathbf{Y}}\in\mathbb{C}^{N_{t}\times N_{t}}$, it may
appear that the complexity of constructing the above bound is $\mathcal{O}(N_{t}^{3})$
due to the computation of $\Bigl(\hat{\mathbf{Y}}_{j}^{(n)}\Bigr)^{-1}$.
However, \emph{we emphasize that this is not the case}. Specifically,
by invoking the matrix-inversion lemma we can write
\begin{align}
\!\!\!\Bigl(\hat{\mathbf{Y}}_{j}^{(n)}\Bigr)^{-1} & =\bigl(\hat{\mathbf{V}}_{j+1}^{(n)H}\hat{\mathbf{V}}_{j+1}^{(n)}+\hat{\mathbf{Y}}_{j+1}^{(n)}\bigr)^{-1}\\
 & =\bigl(\hat{\mathbf{Y}}_{j+1}^{(n)}\bigr)^{-1}-\bigl(\hat{\mathbf{Y}}_{j+1}^{(n)}\bigr)^{-1}\hat{\mathbf{V}}_{j+1}^{(n)H}\bigl(\mathbf{I}+\label{eq:Yj:recursive}\\
 & \quad\hat{\mathbf{V}}_{j+1}^{(n)}\bigl(\hat{\mathbf{Y}}_{j+1}^{(n)}\bigr)^{-1}\hat{\mathbf{V}}_{j+1}^{(n)H}\bigr)^{-1}\hat{\mathbf{V}}_{j+1}^{(n)}\bigl(\hat{\mathbf{Y}}_{j+1}^{(n)}\bigr)^{-1}\\
 & =\bigl(\hat{\mathbf{Y}}_{j+1}^{(n)}\bigr)^{-1}-\bar{\mathbf{B}}_{j+1}^{(n)H}\bigl(\mathbf{I}+\bar{\mathbf{B}}_{j+1}^{(n)}\hat{\mathbf{V}}_{j+1}^{(n)H}\bigr)^{-1}\bar{\mathbf{B}}_{j+1}^{(n)},
\end{align}
for $j=1,2,\ldots,K-1$. The above equation indeed suggests \emph{a
recursive method} to compute $\Bigl(\hat{\mathbf{Y}}_{j}^{(n)}\Bigr)^{-1}$
efficiently. Suppose the inverse of $\hat{\mathbf{Y}}_{j+1}^{(n)}$
is known. Then, we only need to compute the inverse of the matrix
$\mathbf{I}+\bar{\mathbf{B}}_{j+1}^{(n)}\hat{\mathbf{V}}_{j+1}^{(n)H}\in\mathbb{C}^{N_{r}\times N_{r}}$,
which has complexity of $\mathcal{O}(N_{r}^{3})$, to obtain $\Bigl(\hat{\mathbf{Y}}_{j}^{(n)}\Bigr)^{-1}$.
Thus, starting from $\hat{\mathbf{Y}}_{K}^{(n)}=\mathbf{I}$, we can
iteratively compute $\Bigl(\hat{\mathbf{Y}}_{j}^{(n)}\Bigr)^{-1}$
for $j=K-1,K-2,\ldots,1$. In this way, $\mathbf{A}_{j}^{(n)}$ is
also obtained easily.
\end{rem}
Now, using the above lower bound of $h(\mathbf{U})$, we arrive at
the following approximate problem
\begin{equation}
\text{\ensuremath{\underset{\mathbf{U}_{k}}{\maximize}} }\frac{\bar{h}(\mathbf{U};\mathbf{U}^{(n)})}{\sum_{k=1}^{K}\tr(\mathbf{U}_{k}^{H}\mathbf{U}_{k})+N_{t}P_{c}+P_{0}+LNP_{s}}.\label{eq:LB:subproblem}
\end{equation}

\subsubsection{Dinkelbach's Method: Water-filling Solution}

It is important to note that (\ref{eq:LB:subproblem}) is a concave-convex
fractional program since $\bar{h}(\mathbf{U};\mathbf{U}^{(n)})$ is
concave. Thus, Dinkelbach's method can be applied to find the optimal
solution, which leads to the following parameterized problem
\begin{gather}
\underset{\mathbf{U}_{k}}{\maximize}\;f_{\lambda_{D}}(\mathbf{U})=c^{(n)}+\sum\nolimits_{j=1}^{K}2\mathfrak{R}\Bigl(\tr(\mathbf{B}_{j}^{(n)}\mathbf{U}_{j}^{H})\Bigr)\nonumber \\
-\sum\nolimits_{j=1}^{K}\tr\bigl(\mathbf{A}_{j}^{(n)H}\sum\nolimits_{k=j}^{K}\mathbf{H}_{k}^{H}\mathbf{U}_{k}^{H}\mathbf{U}_{k}\mathbf{H}_{k}\bigr)\nonumber \\
-\lambda_{D}\Bigl(\sum\nolimits_{j=1}^{K}\tr(\mathbf{U}_{j}^{H}\mathbf{U}_{j})+N_{t}P_{c}+P_{0}+LNP_{s}\Bigr)\label{eq:dinkelbach:obj}
\end{gather}
where $\lambda_{D}>0$ is a given parameter. Now it is straightforward
to see that the above optimization problem can be solved independently
for each $\mathbf{U}_{j}$, which admits the following closed-form
solution
\begin{equation}
\mathbf{U}_{j}=\mathbf{B}_{j}^{(n)}\Bigl(\mathbf{H}_{j}\sum_{l=1}^{j}\mathbf{A}_{l}^{(n)H}\mathbf{H}_{j}^{H}+\lambda_{D}\mathbf{I}\Bigr)^{-1},\:j=1,\ldots K.\label{eq:Uj_opt}
\end{equation}
Substituting (\ref{eq:Uj_opt}) into (\ref{eq:dinkelbach:obj}) yields
\begin{align}
f_{\lambda_{D}}(\mathbf{U}) & =\sum_{j=1}^{K}\tr\Bigl(\mathbf{B}_{j}^{(n)}\Bigl(\mathbf{H}_{j}\sum_{l=1}^{j}\mathbf{A}_{l}^{(n)H}\mathbf{H}_{j}^{H}+\lambda_{D}\mathbf{I}\Bigr)^{-1}\mathbf{B}_{j}^{(n)H}\Bigr)\nonumber \\
 & \quad+c^{(n)}-\lambda_{D}\Bigl(N_{t}P_{c}+P_{0}+LNP_{s}\Bigr)
\end{align}
Let $\mathbf{H}_{j}\sum_{l=1}^{j}\mathbf{A}_{l}^{(n)H}\mathbf{H}_{j}^{H}=\mathbf{P}_{j}\boldsymbol{\Sigma}_{j}\mathbf{P}_{j}^{H}$
be the eigenvalue decomposition of $\mathbf{H}_{j}\sum_{l=1}^{j}\mathbf{A}_{l}^{(n)H}\mathbf{H}_{j}^{H}$
where $\mathbf{P}_{j}\in\mathbb{C}^{N_{r}\times N_{r}}$ is unitary
and $\boldsymbol{\Sigma}_{j}\in\mathbb{C}^{N_{r}\times N_{r}}$ is
diagonal. Then, we can rewrite $f_{\lambda_{D}}(\mathbf{U})$ as
\begin{gather}
f_{\lambda_{D}}(\mathbf{U})=\sum_{j=1}^{K}\tr\Bigl(\mathbf{B}_{j}^{(n)}\mathbf{P}_{j}^{H}\Bigl(\boldsymbol{\Sigma}_{j}+\lambda_{D}\mathbf{I}\Bigr)^{-1}\mathbf{P}_{j}\mathbf{B}_{j}^{(n)H}\Bigr)\nonumber \\
\quad+c^{(n)}+\lambda_{D}\Bigl(N_{t}P_{c}+P_{0}+LNP_{s}\Bigr)\\
=\sum_{j=1}^{K}\sum_{i=1}^{N_{r}}\frac{d_{ji}}{\sigma_{ji}+\lambda_{D}}+c^{(n)}-\lambda_{D}\Bigl(N_{t}P_{c}+P_{0}+LNP_{s}\Bigr)\label{eq:f_lambdaD}
\end{gather}
where $d_{ji}$ is the $i$th diagonal element of $\mathbf{P}_{j}^{H}\mathbf{B}_{j}^{(n)H}\mathbf{B}_{j}^{(n)}\mathbf{P}_{j}$
and $\sigma_{ji}$ is the $i$th diagonal element of $\boldsymbol{\Sigma}_{j}$.
Next, we need to find $\lambda_{D}$ such that $f_{\lambda_{D}}(\mathbf{U})=0$,
which can be solved using either the bisection method or Newton's
method.

Let $\mathbf{U}_{k}^{\ast}$ be the optimal solution to (\ref{eq:LB:subproblem}),
which is returned by Dinkelbach's method described above. Then $\mathbf{S}_{k}^{\ast}=(\mathbf{U}_{k}^{\ast})\herm\mathbf{U}_{k}^{\ast}$
is the optimal solution to (\ref{eq:Opt_unconst}), for $k=1,\ldots,K$.
Obviously, if $\sum\nolimits_{k=1}^{K}\tr(\mathbf{S}_{k}^{\ast})\le P_{\max}$,
then $\mathbf{S}^{\ast}=(\mathbf{S}_{1}^{\ast},\mathbf{S}_{2}^{\ast},\dots,\mathbf{S}_{K}^{\ast})$
is also optimal to (\ref{eq:Obj_BC}). On the other hand, if $\sum\nolimits_{k=1}^{K}\tr(\mathbf{S}_{k}^{\ast})>P_{\max}$,
it is straightforward to see that the optimal solution to (\ref{eq:Obj_BC})
is obtained by solving the sum-rate maximization problem with the
sum power constraint, which is defined as: 
\begin{subequations}
\label{eq:SRmax:MAC}
\begin{align}
\!\text{\!\!\ensuremath{\underset{\mathbf{S}_{k}}{\maximize}}	} & \ln\left|\mathbf{I}+\sum\nolimits_{k=1}^{K}\mathbf{H}_{k}^{H}\mathbf{S}_{k}\mathbf{H}_{k}\right|\\
\st & \tr(\mathbf{S}_{k})\le P_{\max}.
\end{align}
\end{subequations}
 An efficient method for solving the above problem was proposed in
\cite{perovic2022maximum}, which we omit the details for the sake
of brevity. Let $\hat{\mathbf{S}}=(\hat{\mathbf{S}}_{1},\hat{\mathbf{S}}_{2,}\dots,\hat{\mathbf{S}}_{K})$
be the optimal solution to (\ref{eq:SRmax:MAC}). Then it is easy
to see that the optimal covariance matrices for (\ref{eq:opt_problem_MAC})
are given by
\begin{equation}
\mathbf{S}_{\text{opt}}=\begin{cases}
\mathbf{S}^{\ast} & \sum\nolimits_{k=1}^{K}\tr(\mathbf{S}_{k}^{\ast})\le P_{\max}\\
\hat{\mathbf{S}} & \text{otherwise}.
\end{cases}
\end{equation}
A summary of the described covariance matrix optimization method is
presented in \algref{alg:DPC_cov_mat_opt}.

\subsection{SIM Phase Shift Optimization\label{subsec:SIM-Phase-Shift_DPC}}

Since the total power consumption does not depend on the SIM phase
shifts, the SIM phase shift optimization problem for fixed $\mathbf{S}$
is formulated as 
\begin{subequations}
\begin{align}
\ensuremath{\underset{\boldsymbol{\phi}}{\maximize}}\quad & \kappa(\boldsymbol{\phi})=\ln\left|\mathbf{I}+\sum\nolimits_{k=1}^{K}\mathbf{H}_{k}^{H}\mathbf{S}_{k}\mathbf{H}_{k}\right|\\
\st\quad & \left|\phi_{n}^{l}\right|=1,\forall l,n.
\end{align}
\end{subequations}

Considering the large size of the SIM, we adopt a gradient-based method
to optimize the phase shifts, which consists of the following iterations:
\begin{equation}
\boldsymbol{\phi}^{(n+1)}=\mathcal{P}_{\phi}(\boldsymbol{\phi}^{(n)}+u_{n}\nabla_{\boldsymbol{\phi}}\kappa(\boldsymbol{\phi}^{(n)})),
\end{equation}
where $\mathcal{P}_{\phi}(\cdot)$ denotes the projection operator
that ensures the unit modulus constraint on $\boldsymbol{\phi}$ and
$u_{n}$ is the appropriate step size. The gradient of $\kappa(\boldsymbol{\phi})$
w.r.t. the phase shifts of the SIM is determined by the gradients
of $\kappa(\boldsymbol{\phi})$ w.r.t. the phase shifts of the constituent
SIM layers~as
\begin{align*}
\nabla_{\boldsymbol{\phi}}\kappa(\boldsymbol{\phi}) & =[\nabla_{\boldsymbol{\phi}^{1}}\kappa(\boldsymbol{\phi})^{T}\:\cdots\:\nabla_{\boldsymbol{\phi}^{L}}\kappa(\boldsymbol{\phi})^{T}]^{T}.
\end{align*}
The gradient w.r.t. each $\boldsymbol{\phi}^{L}$ is provided in Theorem
1.\setcounter{thm}{0}
\begin{thm}
The gradient of $h(\boldsymbol{\phi})$ w.r.t. $\boldsymbol{\phi}^{L}$
is given by
\begin{align}
\nabla_{\boldsymbol{\phi}^{l}}\kappa(\boldsymbol{\phi}) & =\vect_{d}(\mathbf{C})\label{eq:grad_phi}
\end{align}
where 
\begin{subequations}
\begin{align}
\mathbf{D} & =\mathbf{I}+\sum\nolimits_{k=1}^{K}\mathbf{H}_{k}^{H}\mathbf{S}_{k}\mathbf{H}_{k}\label{eq:defA}\\
\mathbf{C} & =\boldsymbol{\Theta}^{l+1:L}\sum\nolimits_{k=1}^{K}\ensuremath{\mathbf{G}}_{k}^{H}\mathbf{S}_{k}\mathbf{H}_{k}\mathbf{D}^{-1}\boldsymbol{\Theta}^{1:l-1}(\mathbf{W}^{l})^{H}\label{eq:defC}
\end{align}
\end{subequations}
 where $\boldsymbol{\Theta}^{m:n}=(\mathbf{W}^{m})^{H}(\boldsymbol{\mathbf{\Phi}}^{m})^{H}\cdots(\mathbf{W}^{n})^{H}(\boldsymbol{\mathbf{\Phi}}^{n})^{H}$.
\end{thm}
\begin{IEEEproof}
By differentiating $\kappa(\boldsymbol{\phi})$ w.r.t. $\mathbf{\boldsymbol{\Phi}}^{l}$,
we obtain
\begin{gather}
\!\!\!\!\!\!\text{d}\kappa(\boldsymbol{\phi})=\tr\bigg(\mathbf{D}^{-1}\sum\nolimits_{k=1}^{K}\text{d}(\mathbf{H}_{k}^{H}\mathbf{S}_{k}\mathbf{H}_{k})\bigg)=\nonumber \\
\tr\bigg(\sum\nolimits_{k=1}^{K}\mathbf{S}_{k}\mathbf{H}_{k}\mathbf{D}^{-1}\text{d}\mathbf{H}_{k}^{H}+\sum\nolimits_{k=1}^{K}\mathbf{D}^{-1}\mathbf{H}_{k}^{H}\mathbf{S}_{k}\text{d}\mathbf{H}_{k}\bigg).
\end{gather}
Substituting
\begin{equation}
\text{d}\mathbf{H}_{k}=\ensuremath{\mathbf{G}}_{k}\boldsymbol{\mathbf{\Phi}}^{L}\mathbf{W}^{L}\cdots\text{d}\boldsymbol{\mathbf{\Phi}}^{l}\mathbf{W}^{l}\cdots\boldsymbol{\mathbf{\Phi}}^{1}\mathbf{W}^{1}\label{eq:dHk}
\end{equation}
in the previous equation results in
\begin{equation}
\text{d}\kappa(\boldsymbol{\phi})=\tr(\mathbf{C}\text{d}(\mathbf{\boldsymbol{\Phi}}^{l})^{H}+\mathbf{C}^{H}\text{d}\mathbf{\boldsymbol{\Phi}}^{l})
\end{equation}
where $\mathbf{C}$ is defined in (\ref{eq:defC}). Hence, we have
\begin{equation}
\nabla_{\boldsymbol{\mathbf{\Phi}}^{l}}\kappa(\boldsymbol{\phi})=\mathbf{C}
\end{equation}
and using $\boldsymbol{\mathbf{\Phi}}^{l}=\text{diag}(\boldsymbol{\phi}^{l}),$
we obtain (\ref{eq:grad_phi}).
\end{IEEEproof}
Since all the elements of $\boldsymbol{\phi}$ have unit amplitude,
the projection $\mathcal{P}_{\phi}(\boldsymbol{\phi})$ is defined
by
\begin{equation}
\bar{\phi}_{n}^{l}=\begin{cases}
\phi_{n}^{l}/|\phi_{n}^{l}| & \phi_{n}^{l}\neq0,\\
e^{j\alpha},\alpha\in[0,2\pi] & \phi_{n}^{l}=0.
\end{cases}
\end{equation}

\textcolor{black}{The initial step size $u_{0}$ is chosen as an arbitrary
large positive number. Later, the step size in the $n$-th iteration
of the proposed algorithm $u_{n}$ can be expressed as $u_{0}\rho^{k_{n}}$,
where $k_{n}\ge0$ is the smallest integer that satisfies $\kappa(\boldsymbol{\phi}^{(n+1)})\ge\kappa(\boldsymbol{\phi}^{(n)})+\delta||\boldsymbol{\phi}^{(n+1)}-\boldsymbol{\phi}^{(n)}||^{2}$,
where $\delta>0$ is a small constant and $\rho\in(0,1).$}

Finally, the EE optimization algorithm for a SIM-aided broadcast system
with DPC precoding is given in \algref{alg:DPC_all}. 
\begin{algorithm}[t]
\caption{Proposed algorithm for the EE optimization for a SIM-aided broadcast
system with DPC precoding. \label{alg:DPC_all}}

\SetAlgoNoLine
\DontPrintSemicolon
\LinesNumbered 

\KwIn{$\mathbf{H}$, $\mathbf{S}^{(0)}$, $\boldsymbol{\phi}^{(0)}$,
$\delta>0$, $u_{0}>0$, $\rho\in(0,1)$, $n\leftarrow0$}

\Repeat{convergence of $\eta_{\mathrm{dpc}}$ in (\ref{eq:Obj_BC})
}{

Call \algref{alg:DPC_cov_mat_opt} to obtain $\mathbf{S}^{(n+1)}$\;

\Repeat{$\kappa(\boldsymbol{\phi}^{(n+1)})\ge\kappa(\boldsymbol{\phi}^{(n)})+\delta||\boldsymbol{\phi}^{(n+1)}-\boldsymbol{\phi}^{(n)}||^{2}$}{

$\boldsymbol{\phi}^{(n+1)}=\mathcal{P}_{\phi}(\boldsymbol{\phi}^{(n)}+u_{n}\nabla_{\boldsymbol{\phi}}\kappa(\boldsymbol{\phi}^{(n)}))$\;

\If{$\kappa(\boldsymbol{\phi}^{(n+1)})<\kappa(\boldsymbol{\phi}^{(n)})+\delta||\boldsymbol{\phi}^{(n+1)}-\boldsymbol{\phi}^{(n)}||^{2}$}{

$u_{n}\leftarrow\rho u_{n}$\;

}

}

Calculate $\mathbf{B}$ and $\mathbf{H}$ for $\boldsymbol{\phi}^{(n+1)}$\;

$u_{n+1}\leftarrow u_{n}$\;

$n\leftarrow n+1$\;

}

After $\mathbf{S}_{\text{opt}}$ is obtained, optimal covariance matrices
in the BC are found using the the \acp{BC}-\acp{MAC} duality.
\end{algorithm}

\section{Proposed Solution to LP}

In this section, we propose an optimization method for the EEmax problem
with LP. Similar to the previous section, the precoding matrices are
found in closed-form using Dinkelbach\textquoteright s method, \textcolor{black}{which
also yields a water-filling solution.} The phase shifts for the SIM
meta-elements are optimized by a gradient-based method.

\subsection{Precoding Matrix Optimization}

For fixed $\boldsymbol{\phi}$, the precoding matrix optimization
in (\ref{eq:opt_problem-LP}) reduces to the EEmax problem in conventional
MIMO systems. The structure of our proposed method for precoding matrix
optimization as follows:
\begin{itemize}
\item Reduce the dimension of the precoding matrix optimization problem
by considering an equivalent reformulation.
\item Apply (\ref{eq:quadractic_approx}) to obtain a lower bound of the
achievable rate for LP.
\item Adopt Dinkelbach's method to solve the attained problem, which leads
to water-filling solution.
\end{itemize}

\subsubsection{Problem Reformulation: Dimension Reduction}

It can be observed that the complexity of the direct optimization
of $\mathbf{P}$ is proportional to $N_{t}$, which can cause a significant
complexity burden even for moderate values of $N_{t}$. Consequently,
such a direct optimization method is not practically appealing.

To overcome this issue, we consider an equivalent formulation of (\ref{eq:opt_problem-LP}),
introduced in \cite{zhao2023rethinking}, which reduces the problem's
size and computational complexity. Denoting $\mathbf{H}=[\mathbf{H}_{1}^{T}\:\mathbf{H}_{2}^{T}\:\cdots\:\mathbf{H}_{K}^{T}]^{T}\in\mathbb{C}^{KN_{r}\times N_{t}}$,
(\ref{eq:opt_problem-LP}) can be equivalently written as 
\begin{subequations}
\label{eq:OP-LP-X}
\begin{align}
\!\text{\!\!\ensuremath{\underset{\mathbf{X}}{\maximize}}	} & f(\mathbf{X})=\frac{\sum_{k=1}^{K}\bar{R}_{k}}{\sum_{k=1}^{K}\tr(\bar{\mathbf{H}}\mathbf{X}_{k}\mathbf{X}_{k}^{H})+N_{t}P_{c}+P_{0}+LNP_{s}}\label{eq:Obj_LP_new}\\
\st & \sum\nolimits_{k=1}^{K}\tr(\bar{\mathbf{H}}\mathbf{X}_{k}\mathbf{X}_{k}^{H})\le P_{\max},\label{eq:pow_constr-LP-1}
\end{align}
\end{subequations}
where
\begin{align}
\bar{R}_{k}=\ln\Bigg|\mathbf{I}+\bar{\mathbf{H}}_{k}\mathbf{X}_{k}\mathbf{X}_{k}^{H}\bar{\mathbf{H}}_{k}^{H}\left(\mathbf{I}+\sum\nolimits_{j=1,j\neq k}^{K}\bar{\mathbf{H}}_{k}\mathbf{X}_{j}\mathbf{X}_{j}^{H}\bar{\mathbf{H}}_{k}^{H}\right)^{-1}\Bigg|,\label{eq:Rate-LP}
\end{align}
$\mathbf{X}=[\mathbf{X}_{1}\:\mathbf{X}_{2}\:\cdots\:\mathbf{X}_{K}]\in\mathbb{C}^{KN_{r}\times KN_{r}}$
are new optimization variables, and $\bar{\mathbf{H}}_{k}=\mathbf{H}_{k}\mathbf{H}^{H}\in\mathbb{C}^{N_{r}\times KN_{r}}$
is the $k$-th sub-matrix of $\bar{\mathbf{H}}=\mathbf{H}\mathbf{H}^{H}\in\mathbb{C}^{KN_{r}\times KN_{r}}$.
The equivalence between (\ref{eq:opt_problem-LP}) and (\ref{eq:OP-LP-X})
is a result of \cite[Prop. 2]{zhao2023rethinking}, and the optimal
solutions of the two problems are related as $\mathbf{P}_{k}=\mathbf{H}^{H}\mathbf{X}_{k}$.
When comparing the size of $\mathbf{X}$ and $\mathbf{P}=[\mathbf{P}_{1},\mathbf{P}_{2},\dots,\mathbf{P}_{K}]\in\mathbb{C}^{N_{t}\times KN_{r}}$,
we see that the equivalent formulation in (\ref{eq:OP-LP-X}) significantly
reduces the number of optimization variables, particularly for systems
with large $N_{t}$ (i.e., $N_{t}\gg N_{r}$). Following a similar
approach as in \algref{alg:DPC_cov_mat_opt}, we first derive a solution
for the unconstrained case by dropping (\ref{eq:pow_constr-LP-1}),
with the solution for the constrained case following immediately.

\subsubsection{SCA Method: A Tight Lower Bound of Sum Rate}

Upon close inspection, we observe that the denominator of (\ref{eq:OP-LP-X})
is a quadratic convex function, while the numerator of (\ref{eq:OP-LP-X})
is neither a convex nor concave function. As a result, solving (\ref{eq:OP-LP-X})
is nontrivial. To develop an efficient solution, we again exploit
the inequality, given in (\ref{eq:quadractic_approx}), to obtain
a lower bound of the achievable rate in (\ref{eq:Rate-LP}). Using
the identity $\left|\mathbf{I}+\mathbf{Z}_{k}\mathbf{Z}_{k}^{H}\mathbf{Y}_{k}\right|=\left|\mathbf{I}+\mathbf{Z}_{k}^{H}\mathbf{Y}_{k}\mathbf{Z}_{k}\right|$
to reformulate (\ref{eq:Rate-LP}), the lower bound of the achievable
rate of user $k$ can be expressed as
\begin{align}
\bar{R}_{k} & \ge L_{k}=\ln\left|\mathbf{I}+\hat{\mathbf{Z}}_{k}^{H}\hat{\mathbf{Y}}_{k}^{-1}\hat{\mathbf{Z}}_{k}\right|-\tr(\hat{\mathbf{Z}}_{k}^{H}\hat{\mathbf{Y}}_{k}^{-1}\hat{\mathbf{Z}}_{k})\nonumber \\
 & +2\mathfrak{R}(\tr(\hat{\mathbf{Z}}_{k}^{H}\hat{\mathbf{Y}}_{k}^{-1}\bar{\mathbf{H}}_{k}\mathbf{X}_{k}))\nonumber \\
 & -\sum\nolimits_{j=1}^{K}\tr(\mathbf{X}_{j}^{H}\bar{\mathbf{H}}_{k}^{H}\mathbf{A}_{k}^{H}\bar{\mathbf{H}}_{k}\mathbf{X}_{j})-\tr(\mathbf{A}_{k}^{H})
\end{align}
where $\mathbf{Z}_{k}=\bar{\mathbf{H}}_{k}\mathbf{X}_{k}\in\mathbb{C}^{N_{r}\times N_{r}}$,
$\mathbf{\hat{Z}}_{k}=\bar{\mathbf{H}}_{k}\mathbf{X}_{k}^{(n)}\in\mathbb{C}^{N_{r}\times N_{r}}$,
$\mathbf{Y}_{k}=\sum\nolimits_{j=1,j\neq k}^{K}\bar{\mathbf{H}}_{k}\mathbf{X}_{j}\mathbf{X}_{j}^{H}\bar{\mathbf{H}}_{k}^{H}+\mathbf{I}\in\mathbb{C}^{N_{r}\times N_{r}}$,
$\hat{\mathbf{Y}}_{k}=\sum\nolimits_{j=1,j\neq k}^{K}\bar{\mathbf{H}}_{k}\mathbf{X}_{j}^{(n)}(\mathbf{X}_{j}^{(n)})^{H}\bar{\mathbf{H}}_{k}^{H}+\mathbf{I}\in\mathbb{C}^{N_{r}\times N_{r}}$,
\textcolor{black}{$\mathbf{A}_{k}=\hat{\mathbf{Y}}_{k}^{-1}-(\hat{\mathbf{Z}}_{k}\hat{\mathbf{Z}}_{k}^{H}+\hat{\mathbf{Y}}_{k})^{-1}\in\mathbb{C}^{N_{r}\times N_{r}}$,
and $\mathbf{X}_{k}^{(n)}$ represents the value of $\mathbf{X}_{k}$
at the $n$-th iteration. Consequently,} the resulting approximate
problem of (\ref{eq:OP-LP-X}) is given by 
\begin{align}
\underset{\mathbf{X}}{\maximize} & \bar{f}(\mathbf{X})=\frac{\sum_{k=1}^{K}L_{k}}{\sum_{k=1}^{K}\tr(\bar{\mathbf{H}}\mathbf{X}_{k}\mathbf{X}_{k}^{H})+N_{t}P_{c}+P_{0}+LNP_{s}}\label{eq:Obj_LP_LB}
\end{align}
which is a concave-convex optimization problem.

\subsubsection{Dinkelbach\textquoteright s method: Closed-form Solution}

Next, we apply Dinkelbach\textquoteright s method to solve (\ref{eq:Obj_LP_LB}),
leading to the following optimization problem
\begin{equation}
\underset{\mathbf{X}}{\maximize}\;M_{\lambda_{L}}(\mathbf{X})\label{eq:OP_Dink_X}
\end{equation}
where 
\begin{align}
M_{\lambda_{L}}(\mathbf{X}) & =\sum\nolimits_{k=1}^{K}L_{k}-\lambda_{L}\Big(\sum\nolimits_{k=1}^{K}\tr(\bar{\mathbf{H}}\mathbf{X}_{k}\mathbf{X}_{k}^{H})\nonumber \\
 & +N_{t}P_{c}+P_{0}+LNP_{s}\Big)
\end{align}
and $\lambda_{L}\ge0$ is a given parameter.

Applying the \ac{KKT} first-order optimality condition to (\ref{eq:OP_Dink_X}),
we take the gradient w.r.t. $\mathbf{X}_{j}^{*}$ and set it to \textbf{0},
giving
\begin{algorithm}[t]
\caption{Optimization of the precoding matrices.\label{alg:LP-Xopt}}

\SetAlgoNoLine
\DontPrintSemicolon
\LinesNumbered 

\KwIn{$\mathbf{H}$, $\mathbf{X}^{(0)}$, $\boldsymbol{\phi}$,
$\lambda_{L}^{(0)}\ge0$, $m\leftarrow0$}

\Repeat{$f(\mathbf{X}^{(n+1)})-f(\mathbf{X}^{(n)})>\epsilon_{1,L}$
}{

$n\leftarrow0$\;

\Repeat{ $\lambda_{L}^{(m+1)}-\lambda_{L}^{(m)}>\epsilon_{2,L}$
}{

Calculate all $\mathbf{\hat{Z}}_{k}$, $\mathbf{\hat{Y}}_{k}$, $\mathbf{A}_{k}$\;

Calculate $\mathbf{X}_{\text{opt}}$ according to (\ref{eq:Xj_opt})
or (\ref{eq:Xj_other})\; \label{Step_Xopt}

$\mathbf{X}^{(n+1)}\leftarrow\mathbf{X}_{\text{opt}}$ \;

$n\leftarrow n+1$\;

}

$\lambda_{L}^{(m+1)}\leftarrow f(\mathbf{X}^{(n)})$ according to
(\ref{eq:Obj_LP_new}) \;

$m\leftarrow m+1$\;

$\mathbf{X}^{(0)}\leftarrow\mathbf{X}^{(n)}$\;

}

\eIf{$\sum_{k=1}^{K}\tr(\bar{\mathbf{H}}\mathbf{X}_{k}\mathbf{X}_{k}^{H})\le P_{\max}$}{

$\mathbf{X}_{\text{opt}}=\mathbf{X}$\;}{

$\mathbf{X}_{\text{opt}}$ obtained from \cite{zhao2023rethinking}
\;

}

\For{$k=1$ $\mathbf{to}$ $K$}{

$\mathbf{P}_{k}=\mathbf{H}^{H}\mathbf{X}_{\text{opt},k}$ \;

}
\end{algorithm}
\begin{equation}
(\hat{\mathbf{Z}}_{j}^{H}\hat{\mathbf{Y}}_{j}^{-1}\bar{\mathbf{H}}_{j})^{H}-\sum\nolimits_{k=1}^{K}\bar{\mathbf{H}}_{k}^{H}\mathbf{A}_{k}^{H}\bar{\mathbf{H}}_{k}\mathbf{X}_{j}-\lambda_{L}\bar{\mathbf{H}}\mathbf{X}_{j}=\boldsymbol{0}\label{eq:kkt:Xj}
\end{equation}

To solve for $\mathbf{X}_{j}$ we differentiate two cases. If $\mathbf{H}$
is a row-rank matrix, e.g. (i.e., $KN_{r}<N_{t}$), then $\bar{\mathbf{H}}$
is invertible, and thus (\ref{eq:kkt:Xj}) results in
\begin{align}
\mathbf{X}_{j} & =\left(\sum\nolimits_{k=1}^{K}\bar{\mathbf{H}}_{k}^{H}\mathbf{A}_{k}^{H}\bar{\mathbf{H}}_{k}+\lambda_{L}\bar{\mathbf{H}}\right)^{-1}\bar{\mathbf{H}}_{j}^{H}\hat{\mathbf{Y}}_{j}^{-1}\hat{\mathbf{Z}}_{j}.\label{eq:Xj_opt}
\end{align}
If $\mathbf{H}$ is column-rank matrix (i.e., $KN_{r}>N_{t}$), we
rewrite (\ref{eq:kkt:Xj}) as
\begin{equation}
\mathbf{H}\Bigl(\mathbf{H}_{j}^{H}\hat{\mathbf{Y}}_{j}^{-1}\hat{\mathbf{Z}}_{j}-\sum\nolimits_{k=1}^{K}\mathbf{H}_{k}^{H}\mathbf{A}_{k}^{H}\bar{\mathbf{H}}_{k}\mathbf{X}_{j}-\lambda_{L}\mathbf{H}^{H}\mathbf{X}_{j}\Bigr)=\boldsymbol{0}
\end{equation}
and finally obtain
\begin{equation}
\mathbf{X}_{j}=\Bigl(\sum\nolimits_{k=1}^{K}\mathbf{H}_{k}^{H}\mathbf{A}_{k}^{H}\bar{\mathbf{H}}_{k}+\lambda_{L}\mathbf{H}^{H}\Bigr)^{+}\mathbf{H}_{j}^{H}\hat{\mathbf{Y}}_{j}^{-1}\hat{\mathbf{Z}}_{j}.\label{eq:Xj_other}
\end{equation}

If the solution obtained from (\ref{eq:Xj_opt}) or (\ref{eq:Xj_other})
satisfies the power constraint (\ref{eq:pow_constr-LP-1}), then it
is also the optimal solution to (\ref{eq:OP-LP-X}). Otherwise, the
optimal $\mathbf{X}$ is determined by solving the achievable sum
rate optimization problem
\begin{subequations}
\begin{align}
\!\text{\!\!\ensuremath{\underset{\mathbf{X}}{\maximize}}	} & \sum\nolimits_{k=1}^{K}\bar{R}_{k}\label{eq:Obj_LP_new-1}\\
\st & \sum\nolimits_{k=1}^{K}\tr(\bar{\mathbf{H}}\mathbf{X}_{k}\mathbf{X}_{k}^{H})\le P_{\max},\label{eq:pow_constr-LP-1-2}
\end{align}
\end{subequations}
which can be solved by \cite[Algorithm 1]{zhao2023rethinking}. The
described optimization algorithm is summarized in \algref{alg:LP-Xopt}. 

\subsection{SIM Phase Shift Optimization}

For a fixed $\mathbf{P}$, the SIM phase shift optimization problem
is formulated as 
\begin{subequations}
\begin{align}
\ensuremath{\underset{\boldsymbol{\phi}}{\maximize}}\quad & \tau(\boldsymbol{\phi})=\sum\nolimits_{k=1}^{K}R_{\text{L},k}(\boldsymbol{\phi})\\
\st\quad & \left|\phi_{n}^{l}\right|=1,\forall l,n,
\end{align}
\end{subequations}
 where we have used the rate expression in (\ref{eq:Rate_user}) instead
of (\ref{eq:Rate-LP}). This choice not only simplifies the gradient
computation of the objective function w.r.t $\boldsymbol{\phi}$ but
also leads to faster convergence rate as explained in \cite{perovic2021achievable}.

The optimization of the phase shifts of the BS SIM follows the same
steps as in the case of DPC discussed in subsection \ref{subsec:SIM-Phase-Shift_DPC}.
Specifically, the phase shifts are iteratively updated according to
\begin{algorithm}[t]
{\small\caption{Proposed algorithm for the EE optimization for a SIM-aided broadcast
system with LP. \label{alg:LP_whole_alg}}

\SetAlgoNoLine
\DontPrintSemicolon
\LinesNumbered 

\KwIn{$\mathbf{H}$, $\mathbf{X}^{(0)}$, $\boldsymbol{\phi}^{(0)}$,
$\lambda^{(0)}\ge0$, $\delta>0$, $t_{n}>0$, $\rho\in(0,1)$, $n\leftarrow0$}

\Repeat{convergence of $\eta_{\mathrm{lp}}$ in (\ref{eq:Obj_LP})
}{

Call \algref{alg:LP-Xopt} to obtain $\mathbf{P}^{(n+1)}$\;

\Repeat{ $\tau(\boldsymbol{\phi}^{(n+1)})\ge\tau(\boldsymbol{\phi}^{(n)})+\delta||\boldsymbol{\phi}^{(n+1)}-\boldsymbol{\phi}^{(n)}||^{2}$
}{

$\boldsymbol{\phi}^{(n+1)}=\mathcal{P}_{\phi}(\boldsymbol{\phi}^{(n)}+t_{n}\nabla_{\boldsymbol{\phi}}\tau(\boldsymbol{\phi}^{(n)}))$\;

\If{ $\tau(\boldsymbol{\phi}^{(n+1)})<\tau(\boldsymbol{\phi}^{(n)})+\delta||\boldsymbol{\phi}^{(n+1)}-\boldsymbol{\phi}^{(n)}||^{2}$
}{

$t_{n}\leftarrow\rho t_{n}$\;

}

}

$n\leftarrow n+1$\;

}}
\end{algorithm}
\begin{equation}
\boldsymbol{\phi}^{(n+1)}=\mathcal{P}_{\phi}(\boldsymbol{\phi}^{(n)}+t_{n}\nabla_{\boldsymbol{\phi}}\tau(\boldsymbol{\phi}^{(n)})),
\end{equation}
where $t_{n}$ is the appropriate step size. Also, the gradient $\nabla_{\boldsymbol{\phi}}\tau(\boldsymbol{\phi})$
is given by $\nabla_{\boldsymbol{\phi}}\tau(\boldsymbol{\phi})=[\nabla_{\boldsymbol{\phi}^{1}}\tau(\boldsymbol{\phi})^{T}\:\cdots\:\nabla_{\boldsymbol{\phi}^{L}}\tau(\boldsymbol{\phi}))^{T}]^{T}$,
where $\nabla_{\boldsymbol{\phi}^{l}}\tau(\boldsymbol{\phi})$ is
expressed in the following theorem.
\begin{thm}
\label{thm:grad:phase:shift}The gradients of $\tau(\boldsymbol{\phi})$
w.r.t. the l-th layer of the SIM at the BS is given by
\begin{gather}
\nabla_{\boldsymbol{\phi}^{l}}\tau(\boldsymbol{\phi})=\vect_{d}\left(\sum\nolimits_{k=1}^{K}\boldsymbol{\Theta}^{l+1:L}\mathbf{G}_{k}^{H}\mathbf{F}_{1,k}^{-1}\mathbf{H}_{k}\hat{\mathbf{P}}_{s}\boldsymbol{\Theta}^{1:l-1}\mathbf{W}_{l}^{H}\right)\nonumber \\
-\vect_{d}\left(_{k=1}^{K}\sum\nolimits_{k=1}^{K}\boldsymbol{\Theta}^{l+1:L}\mathbf{G}_{k}^{H}\mathbf{F}_{2,k}^{-1}\mathbf{H}_{k}\hat{\mathbf{P}}_{k}\boldsymbol{\Theta}^{1:l-1}\mathbf{W}_{l}^{H}\right)\label{eq:phi_grad}
\end{gather}
where $\boldsymbol{\Theta}^{m:n}=(\mathbf{W}^{m})^{H}(\boldsymbol{\mathbf{\Phi}}^{m})^{H}\cdots(\mathbf{W}^{n})^{H}(\boldsymbol{\mathbf{\Phi}}^{n})^{H}$,
$\hat{\mathbf{P}}_{s}=\sum_{j=1}^{K}\mathbf{P}_{j}\mathbf{P}_{j}^{H}$,
$\hat{\mathbf{P}}_{k}=\sum_{j=1,j\neq k}^{K}\mathbf{P}_{j}\mathbf{P}_{j}^{H}$,
$\mathbf{F}_{1,k}=\mathbf{I}+\mathbf{H}_{k}\hat{\mathbf{P}}_{s}\mathbf{H}_{k}^{H}$
and $\mathbf{F}_{2,k}=\mathbf{I}+\mathbf{H}_{k}\hat{\mathbf{P}}_{k}\mathbf{H}_{k}^{H}$.
\end{thm}
\begin{IEEEproof}
See the Appendix.
\end{IEEEproof}
The outline of the proposed algorithm is given in \algref{alg:LP_whole_alg}. 

\section{Computational Complexity}

In this section, the computational complexity for SIM-aided broadcast
systems with DPC and LP is obtained by counting the required number
of complex multiplications. In the following complexity analysis,
for simplicity, we assume that $N,N_{t}\gg N_{r}$ which is a typical
case for a SIM-based broadcast communication system.

\subsection{DPC Precoding}

The covariance matrix optimization is performed by \algref{alg:DPC_cov_mat_opt},
with the complexity of obtaining $\mathbf{U}$ from $\mathbf{S}$
being negligible. In addition, the complexity of one iteration of
inner loop (i.e., lines 2 to 5 in \algref{alg:DPC_cov_mat_opt}) is
$\mathcal{O}(N_{t}^{2}N_{r})$. Also, the complexity of line 6 is
approximately $\mathcal{O}(KN_{t}N_{r}^{2})$, while the complexity
of line 7 can be neglected. Let $I_{U}$ denote the number of inner
loops (i.e., lines 1 to 1 in \algref{alg:DPC_cov_mat_opt}), then
the complexity of \algref{alg:DPC_cov_mat_opt} is given by $\mathcal{O}(I_{U}KN_{t}^{2}N_{r})$.

Optimizing the SIM phase shifts requires $\mathcal{O}(KNN_{t}N_{r})$
multiplications for computing $\sum\nolimits_{k=1}^{K}\ensuremath{\mathbf{G}}_{k}^{H}\mathbf{S}_{k}\mathbf{H}_{k}$.
The matrix inversion $\mathbf{A}^{-1}$ and its multiplication contribute
$\mathcal{O}(N_{t}^{3}+NN_{t}^{2})$ multiplications. Since $(\mathbf{W}^{m})^{H}(\boldsymbol{\mathbf{\Phi}}^{m})^{H}$
is \emph{precomputed} and only the diagonal elements are needed in
(\ref{eq:grad_phi}), the additional complexity is $\mathcal{O}(LN^{3})$.
Hence, the complexity of the gradient calculation is $\mathcal{O}(NN_{t}^{2}+LN^{3})$.
After obtaining $\boldsymbol{\phi}^{(n+1)}$, the complexity of calculating
$\mathbf{B}$ and $\mathbf{H}$ is $\mathcal{O}(LN^{3})$. In addition,
$\mathcal{O}(KN_{t}N_{r}(N_{t}+N_{r})+N_{t}^{3})\approx\mathcal{O}(N_{t}^{3})$
multiplications are needed to obtain $h(\boldsymbol{\phi}^{(n+1)})$.
The complexity of the SIM phase shifts optimization is $\mathcal{O}(NN_{t}^{2}+I_{\phi,D}(LN^{3}+N_{t}^{3}))$,
where $I_{\phi,D}$ is the number of line search iterations.

Therefore, the overall computational complexity for single iteration
of \algref{alg:DPC_all} is given by
\begin{align}
\!\!\!\!\!\!C_{\text{SIM-DPC}} & =\mathcal{O}(I_{U}KN_{t}^{2}N_{r}+NN_{t}^{2}+I_{\phi,D}(LN^{3}+N_{t}^{3})).
\end{align}

\subsection{Linear Precoding (LP)}

The optimization of the precoding matrices follows \algref{alg:LP-Xopt},
where computing $\bar{\mathbf{H}}$ requires $\mathcal{O}(K^{2}N_{t}N_{r}^{2})$
multiplications. Computing the initial $\mathbf{X}$ from $\mathbf{P}$
requires $\mathcal{O}(K^{2}N_{t}N_{r}^{2}+K^{3}N_{r}^{3})$ multiplications.
Furthermore, the complexity of calculating all $\mathbf{\hat{Z}}_{k}$,
$\mathbf{\hat{Y}}_{k}$ and $\mathbf{A}_{k}$ is $\mathcal{O}(K^{3}N_{r}^{3}+K^{2}N_{r}^{3})$.
To determine $\mathbf{X}$,
\begin{itemize}
\item if $N_{t}\ge KN_{r}$, solving (\ref{eq:Xj_opt}) requires $\mathcal{O}(K^{3}N_{r}^{3})$
multiplications, or otherwise according to
\item if $N_{t}<KN_{r}$, solving (\ref{eq:Xj_other}) requires $\mathcal{O}(K^{3}N_{r}^{3}+K^{3}N_{t}N_{r}^{2})\approx\mathcal{O}(K^{3}N_{t}N_{r}^{2})$
multiplications.
\end{itemize}
Thus, the computational complexity of precoding matrix optimization
is written as{\small
\begin{equation}
C_{x}=\begin{cases}
\mathcal{O}(K^{3}N_{r}^{3}) & N_{t}\ge KN_{r}\\
\mathcal{O}(K^{3}N_{t}N_{r}^{2}) & N_{t}<KN_{r}.
\end{cases}\label{eq:Comp_lin_prec_X}
\end{equation}
}The complexity for calculating $\sum_{k}\bar{R}_{k}$ and $f(\mathbf{X}^{(n)})$
can be neglected. Let $I_{X}$ denote the number of inner loops (i.e.,
lines 3 to 8 in \algref{alg:LP-Xopt}) , then the complexity of \algref{alg:LP-Xopt}
is given by $\mathcal{O}(I_{X}C_{x})$.

To optimize the phase shifts of the SIM, we need $\mathcal{O}(KN_{t}^{2})$
multiplications to obtain $\hat{\mathbf{P}}_{s}$ and all $\hat{\mathbf{P}}_{k}$
matrices, while calculating $\mathbf{F}_{1,k}$ and $\mathbf{F}_{2,k}$
requires $\mathcal{O}(KN_{t}N_{r}(N_{t}+N_{r}))$ multiplications,
which also holds for $\mathbf{F}_{1,k}^{-1}\mathbf{H}_{k}\hat{\mathbf{P}}_{s}-\mathbf{F}_{2,k}^{-1}\mathbf{H}_{k}\hat{\mathbf{P}}_{k}$.
Multiplying these terms with $\ensuremath{\mathbf{G}}_{k}^{H}$ adds
$\mathcal{O}(KNN_{t}^{2})$ multiplications. As mentioned earlier,
since $(\mathbf{W}^{m})^{H}(\boldsymbol{\mathbf{\Phi}}^{m})^{H}$
are precomputed and only the diagonal elements are need in (\ref{eq:phi_grad}),
the additional complexity is $\mathcal{O}(LN^{3})$. Hence, the complexity
of the gradient calculation is $\mathcal{O}(KNN_{t}^{2}+LN^{3})$.
After obtaining $\boldsymbol{\phi}^{(n+1)}$, the complexity of calculating
$\mathbf{B}$ and $\mathbf{H}$ is $\mathcal{O}(LN^{3})$. To obtain
all terms $\sum\nolimits_{j}\mathbf{H}_{k}\mathbf{P}_{j}\mathbf{P}_{j}^{H}\mathbf{H}_{k}^{H},$
we need $\mathcal{O}(KN_{t}N_{r}(N_{t}+N_{r}))$ multiplications.
Any additional complexity for computing $g(\boldsymbol{\phi}^{(n+1)})$
can be neglected. Hence, the complexity of optimizing the SIM phase
shifts is $\mathcal{O}(KNN_{t}^{2}+I_{\phi,L}LN^{3})$, where $I_{\phi,L}$
is the number of line search steps.

Therefore, the overall computational complexity for one iteration
of \algref{alg:LP_whole_alg} is given by
\begin{equation}
C_{\text{SIM-LP}}=\mathcal{O}(I_{X}C_{x}+KNN_{t}^{2}+I_{\phi,L}LN^{3}).
\end{equation}

\section{Convergence}

We now prove the convergence of \algref{alg:DPC_all}. First, for
a given $\boldsymbol{\phi}$, \algref{alg:DPC_all} achieves monotonic
convergence, which can be shown as follows. From (\ref{eq:LB:inequality}),
it holds that
\begin{gather}
g(\mathbf{U}^{(n+1)})=\frac{h(\mathbf{U}^{(n+1)})}{\sum_{k=1}^{K}\tr(\mathbf{U}_{k}^{(n+1)H}\mathbf{U}_{k}^{(n+1)})+N_{t}P_{c}+P_{0}+LNP_{s}}\\
\stackrel{\textrm{(a)}}{\geq}\frac{\bar{h}(\mathbf{U}^{(n+1)};\mathbf{U}^{(n)})}{\sum_{k=1}^{K}\tr(\mathbf{U}_{k}^{(n+1)H}\mathbf{U}_{k}^{(n+1)})+N_{t}P_{c}+P_{0}+LNP_{s}}\\
\stackrel{\textrm{(b)}}{\geq}\frac{\bar{h}(\mathbf{U}^{(n)};\mathbf{U}^{(n)})}{\sum_{k=1}^{K}\tr(\mathbf{U}_{k}^{(n)H}\mathbf{U}_{k}^{(n)})+N_{t}P_{c}+P_{0}+LNP_{s}}\\
\stackrel{\textrm{(c)}}{\geq}\frac{h(\mathbf{U}^{(n)})}{\sum_{k=1}^{K}\tr(\mathbf{U}_{k}^{(n)H}\mathbf{U}_{k}^{(n)})+N_{t}P_{c}+P_{0}+LNP_{s}}=g(\mathbf{U}^{(n)})
\end{gather}
where $\textrm{(a)}$ follows from (\ref{eq:LB:inequality}), $\textrm{(b)}$
holds because $\mathbf{U}^{(n+1)}$ is the optimal solution to (\ref{eq:LB:subproblem})
and that the optimal objective is no less than the objective of a
feasible point, and $\textrm{(c)}$ is true because it is easy to
check that $\bar{h}(\mathbf{U}^{(n)};\mathbf{U}^{(n)})=h(\mathbf{U}^{(n)})$,
i.e. the equality in (\ref{eq:LB:inequality}) occurs when $\mathbf{U}=\mathbf{U}^{(n)}$.
Regarding $\textrm{(b)}$ above, note again that since (\ref{eq:LB:subproblem})
is a concave-convex fractional program, Dinkelbach's method is guaranteed
to converge to its optimal solution. Consequently, the sequence $\{g(\mathbf{U}^{(n)})\}$
increases monotonically to an optimal solution to (\ref{eq:Opt_unconst-precoder}),
and thus, \algref{alg:DPC_all} is able to compute an optimal solution
to (\ref{eq:opt_problem_MAC}). Next, for given covariance matrices,
the SIM phase shift is optimized by a standard projected gradient
method, for which the convergence is guaranteed. Also, the projected
gradient method always yields an improved solution. In other words,
\algref{alg:DPC_all} generates a non-decreasing objective sequence.
Since the feasible sets for the covariance matrices and phase shifts
are continuous, the objective sequence produced by \algref{alg:DPC_all}
is guaranteed to converge.
\begin{figure}[t]
\centering{}\includegraphics[width=8.85cm]{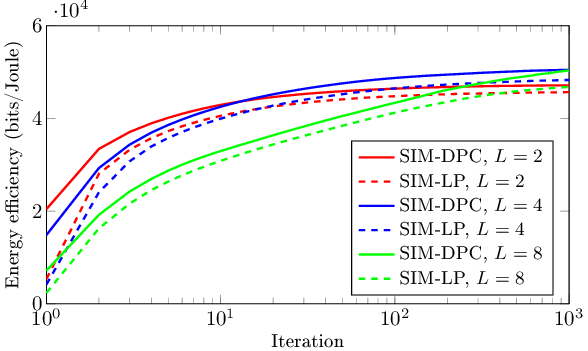}\caption{Convergence of the proposed algorithms for different number of SIM
layers.\label{fig:Convergence}}
\end{figure}

Since \algref{alg:LP_whole_alg} uses a similar method for the optimization
of the precoding matrices, as \algref{alg:DPC_all} for the optimization
of the covariance matrices, we can prove, following the same derivation
steps, that for a given $\boldsymbol{\phi}$ \algref{alg:LP_whole_alg}
is guaranteed to provide an optimal solution to (\ref{eq:LB:subproblem}).
In addition, a gradient-based optimization of the SIM phase shifts
always increase the objective function. Moreover, the feasible sets
for the precoding matrices and phase shifts are continuous, which
ensures the convergence of the objective sequence in \algref{alg:LP_whole_alg}.

\section{Simulation Results \label{sec:Sim-Res}}

In this section, we evaluate the EE of the proposed schemes with DPC
and LP, which are denoted as SIM-DPC and SIM-LP, respectively, through
Monte Carlo simulations and compare them with three benchmark schemes:
\begin{enumerate}
\item \emph{LP-NoSIM}: The first benchmark scheme uses LP without SIM integration,
with a total power consumption of $P_{t}+N_{t}P_{c}+P_{0}.$
\item \emph{SIM-NoLP}: The second scheme, omits digital precoding, feeding
data streams directly to transmit antennas while optimizing SIM phase
shifts as in \algref{alg:LP_whole_alg}.\textcolor{black}{{} Let $N_{t,k}$
BS antennas are allocated for transmission to the $k$-th user. Since
all $N_{t}$ transmit antennas are active, $N_{t,k}$ equals $\left\lfloor N_{t}/K\right\rfloor $
or $\left\lfloor N_{t}/K\right\rfloor +1$. Thus, each data stream
of user $k$ is transmitted by $\left\lfloor N_{t,k}/N_{r}\right\rfloor $
or $\left\lfloor N_{t,k}/N_{r}\right\rfloor +1$ BS antennas.}
\item \emph{SIM-NoLP-RedRF}: The third scheme excludes LP but differs from
SIM-NoLP by utilizing a reduced number of transmit antennas (i.e.,
RF chains). Specifically, it activates $KN_{r}$ transmit antennas
(i.e., RF chains), each transmitting an independent data stream, while
the remaining transmit antennas remain inactive. As such, the total
power consumption of \textbf{$P_{t}+KN_{r}P_{c}+P_{0}+LNP_{s}$}.
\textcolor{black}{Note that for the case when $KN_{r}>N_{t}$, SIM-NoLP
transmits signal to a subset of randomly chosen users, while LP-NoSIM-RedRF
cannot operate. To ensure fair comparison, data for SIM-NoLP and LP-NoSIM-RedRF
are scaled by $\sqrt{P_{\max}/N_{t}}$ and $\sqrt{P_{\max}/KN_{r}},$
respectively.}
\end{enumerate}

\paragraph*{Channel Modeling}

The channel matrix between the BS and user $k$ is modeled as $\mathbf{G}_{k}=\bar{\mathbf{G}}_{k}\mathbf{R}_{\text{T}}^{1/2}\in\mathbb{C}^{N_{r}\times L}$,
where $\bar{\mathbf{G}}_{k}\in\mathbb{C}^{N_{r}\times L}$ denotes
the channel between the last SIM layer and the receiver, distributed
as $\mathcal{CN}(0,\beta\mathbf{I})$. The path-loss between the BS
and user $k$ is given by $\beta(d_{k})=\beta(d_{0})+10b\log_{10}(d_{k}/d_{0})$,
where $\beta(d_{0})=20\log_{10}(4\pi d_{0}/\lambda)$ is the free
space path-loss at the reference distance $d_{0}$, $b$ is the path-loss
exponent, and $d_{k}$ is the distance between the BS and user $k$.
Moreover, $\mathbf{R}_{\text{T}}\in\mathbb{C}^{L\times L}$ is the
spatial correlation matrix of the SIM, defined as in \cite[Eq. (14), (15)]{an2023stackedholo}.

\paragraph*{System Parameters}

In the simulation setup, parameters are set as follows: $\lambda=5\,\text{cm}$
(i.e., $f=6\,\text{GHz}$), $N_{t}=16$, $N_{r}=2$, $W=100\thinspace\text{kHz}$,
$b=3.5$, $d_{0}=1\,\text{m}$, $L=4$, $K=4$, \textcolor{black}{$P_{\max}=5\,\mathrm{W}$}\textcolor{blue}{{}
}and $\sigma^{2}=-110\,\text{dB}$. Unless specified otherwise, the
number of meta-elements per SIM layer, $N$, is 100. The BS antennas
are placed in a planar array parallel to the $xy$-plane, with the
midpoint at $(30\,\mathrm{m},0,0)$ and inter-antenna spacing of $\lambda/2$.
SIM layers are also placed parallel to the $xy$-plane, with the midpoint
of the $l$-th layer at $(30\,\mathrm{m},0,l\lambda/2)$. Meta-elements,
each with dimension $\lambda/2\times\lambda/2$, are uniformly placed
in a square grid on each SIM layer. \textcolor{black}{Hence, any mutual
coupling between different meta-elements can be neglected.} Users\textquoteright{}
ULAs are parallel to the $x$-axis, with the midpoint of the $k$-th
user\textquoteright s ULA at $(x_{k},y_{k},z_{k})$. Users\textquoteright{}
coordinates are uniformly distributed such that $\ensuremath{x_{k}\in[1.6\,\text{m},2\,\text{m}]},\ensuremath{y_{k}\in[-20\,\text{m},20\,\text{m}]},\ensuremath{z_{k}\in[80\,\text{m},120\,\text{m}]}$.
The circuit power per RF chain is $P_{c}=30\text{\thinspace dBm},$
the basic power consumption at the BS is $P_{0}=40\thinspace\text{dBm}$
\cite{you2022energy,he2013coordinated}, and the power consumption
per SIM meta-element is $P_{s}=10\thinspace\text{dBm}$ \cite{wang2024reconfigurable,huang2019reconfigurable}.
Optimization parameters are set $\epsilon_{D}=10^{-6}$ for DPC, and
$\epsilon_{1,L}=\epsilon_{2,L}=10^{-6}$ for LP. Gradient-based methods
use an initial step size of 1000, $\rho=1/2$ and $\delta=10^{-3}$.
Results are averaged over 200 independent channel realizations.

\paragraph*{\textcolor{black}{Initialization}}

\textcolor{black}{For DPC the initial $\mathbf{S}_{k}$ matrices are
randomly generated Hermitian matrices satisfying (\ref{eq:pow_constr_MAC})
with equality. For LP the initial $\mathbf{X}_{k}$ are randomly generated
matrices satisfying (\ref{eq:pow_constr-LP-1}) with equality. The
initial elements of $\mathbf{\boldsymbol{\phi}}$ are randomly chosen
from the unit circle.}
\begin{figure}[t]
\centering{}\includegraphics[width=8.85cm]{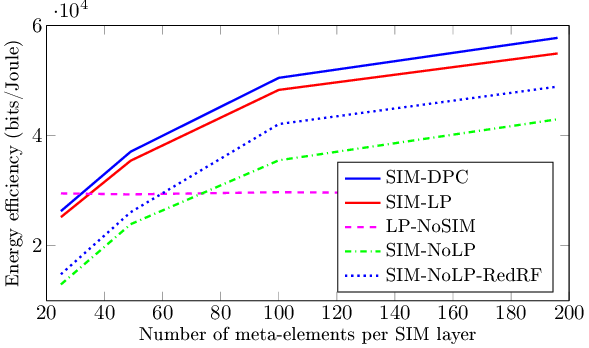}\caption{EE versus number meta-elements per SIM layer. \label{fig:EE_diff_num_metaatoms}}
\end{figure}

The convergence of the proposed algorithms for different numbers of
SIM layers is shown in Fig. \ref{fig:Convergence}. In general, we
can see that the number of iterations required to converge increases
with the number of SIM layers (i.e., meta-elements), consistent with
previous observations in \cite{perovic2024mutual}, where a layer-by-layer
approach for SIM optimization was used. \textcolor{black}{Notably,
the gap in EE performance between SIM-DPC and SIM-LP remains small
for any number of SIM layers, }\textcolor{black}{\emph{making SIM-LP
a practical and efficient solution for SIM-based systems}}\textcolor{black}{.
Moreover, the EE is not monotonic with $L$. For SIM-DPC, the EE at
convergence is almost the same for $L=4$ and $L=8$, while for SIM-LP,
it is slightly higher when $L=8$.} This behavior results from the
fact that the total power consumption scales linearly with the number
of meta-elements, while the achievable sum-rate tends to be saturated
\cite[Fig. 4]{papazafeiropoulos2024achievable}.

Fig. \ref{fig:EE_diff_num_metaatoms} shows the EE for different number
meta-elements per SIM layer ($N=25,49,100,196)$. The EE of both proposed
and the non-precoding benchmark schemes (i.e., SIM-NoLP and SIM-NoLP-RedRF)
increases with $N$ but the increase rate gradually declines due to
the logarithmic behavior of the achievable sum-rate at near-full transmit
power. Among the non-precoding benchmarks, SIM-NoLP-RedRF achieves
higher EE than SIM-NoLP due to reduced power consumption from inactive
RF chains. Moreover, SIM-NoLP\textcolor{black}{{} suffers from increased
multi-stream interference as in this case each data stream is transmitted
from two different transmit antennas without digital precoding, which
potentially worsens as $N$ increases, further enlarging the EE gap
between these benchmarks. On the other hand, the saturation of the
achievable sum-rate, and consequently the EE, with the further increase
of the number of meta-elements results in almost the same enlargement
of the EE for the two benchmark schemes.} For a small $N$, LP-NoSIM
achieves the largest EE, but is eventually inferior as other schemes
benefit from an increased number of meta-elements. Finally, SIM-DPC
achieves slightly higher EE than SIM-LP, with the gap increasing as
$N$ grows.
\begin{figure}[t]
\centering{}\includegraphics[scale=0.75]{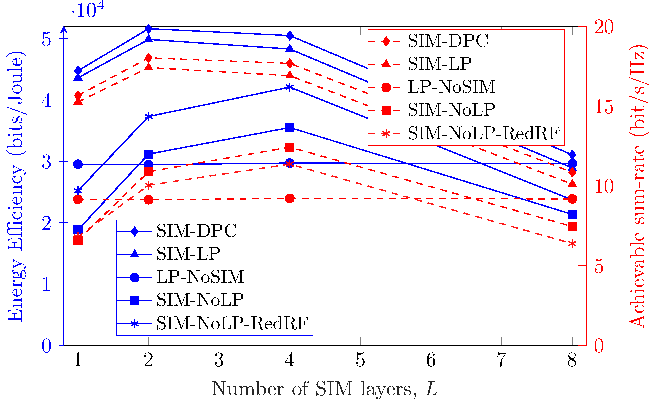}\caption{EE (blue solid lines) and achievable sum-rate (red dashed lines) versus
number SIM layer for the constant number of 400 meta-elements. \label{fig:EE_vs_no_layers}}
\end{figure}

Fig. \ref{fig:EE_vs_no_layers} plots the EE and the achievable sum-rate
of the considered system versus the number of SIM layers, while keeping
a total of $400$ meta-elements.\textcolor{blue}{{} }\textcolor{black}{Importantly,
we find that }\textcolor{black}{\emph{SIM-LP is as good as SIM-DPC
for both performance metrics}}\textcolor{black}{, regardless of the
distribution of meta-elements on SIM layers.} and \emph{neither metric
is monotonic with $L$}. Specifically, both increase from $L=1$ to
$L=2$, which is apparently due to the enhanced beamforming capabilities
offered by multi-layer structures. However, with $L$ increases further,
both performance metrics significantly decrease, potentially approaching
LP-NoSIM. This occurs because fewer meta-elements per layer reduce
the beamforming capability of each layer, while inter-layer propagation
introduces additional path-loss as the number of SIM layers increases.
A similar trend is observed in non-precoding benchmarks (i.e., SIM-NoLP
and SIM-NoLP-RedRF), where the EE and the achievable sum-rate also
decrease for $L>4$. Among them, SIM-NoLP-RedRF achieves higher \textcolor{black}{EE,
despite a lower achievable sum-rate as it operates with fewer RF chains.}
\begin{figure}[t]
\centering{}\includegraphics[width=8.85cm]{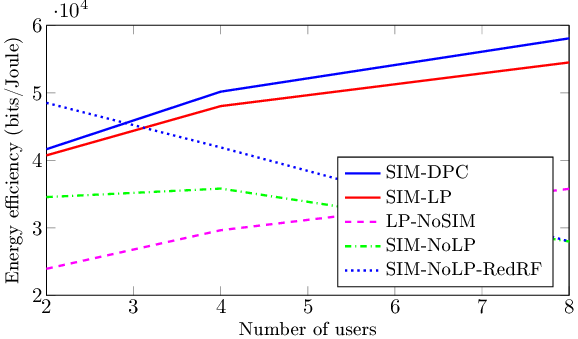}\caption{The EE versus the number users. \label{fig:EE_vs_K}}
\end{figure}

Fig. \ref{fig:EE_vs_K} demonstrates the EE for different number of
users, $K$. For a small $K$, SIM-NoLP-RedRF provides the highest
EE due to its reduced RF chain (and thus power) usage. However, as
$K$ increases, its EE declines since the number of RF chains becomes
comparable to other schemes, and the lack of amplitude control of
the transmitted signal limits its multi-user interference suppression
capability \cite{perovic2022maximum}.\textcolor{blue}{{} }\textcolor{black}{On
the other hand, the EE of SIM-NoLP shows a non-monotonic behavior
with $K$. Specifically, its EE slightly improves from $K=2$ to $K=4$
due to the SIM being able to exploit multi-user diversity gains. However,
when $K$ further increases, the EE of both SIM-NoLP and SIM-NoLP-RedRF
declines. The EE of the proposed SIM-based schemes and LP-NoSIM increase
with $K$ since digital precoding can effectively mitigate the multi-user
interference, allowing them to exploit the multi-user diversity. Moreover,
SIM-DPC achieves slightly higher EE than \mbox{SIM-LP}, and the
gap widens as $K$ increases. This observation further supports LP
as an efficient, practical solution for SIM-based systems.} Overall,
we conclude that integrating SIM with digital precoding consistently
yield the best EE, except for very small~$K$.

\begin{figure}[t]
\centering{}\includegraphics[width=8.85cm]{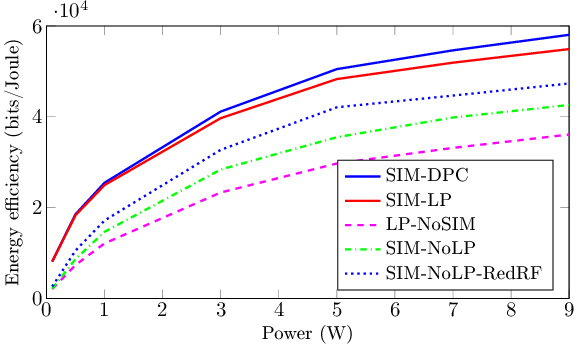}\caption{EE versus the maximum transmit power. \label{fig:EE_vs_Pmax}}
\end{figure}

Next, Fig. \ref{fig:EE_vs_Pmax} illustrates how the EE varies with
the maximum transmit power. All schemes in consideration exhibit a
logarithmic EE increase due to the logarithmic growth of the achievable
rate. \textcolor{black}{At low transmit power, the proposed schemes
(i.e. SIM-DPC and SIM-LP) achieve the same EE, as do the benchmark
schemes. As the transmit power increases,} SIM-DPC consistently shows
a more noticeable improvement in EE over all other schemes, which
is attributed to the superior interference suppression capabilities
of DPC. As expected, the proposed schemes outperform benchmark schemes,
with, SIM-NoNP-RedRF obtaining the highest EE among them because of
fewer RF chains being used. The difference in the EE between SIM-LP
and other benchmark schemes increases with transmit power stabilizes
at higher transmit power levels. 

\begin{figure}[t]
\centering{}\includegraphics[width=8.85cm]{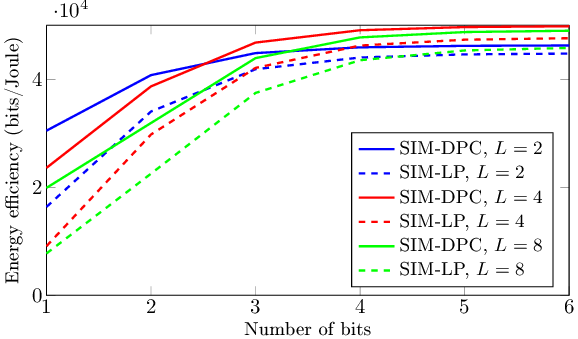}\caption{EE for the case of discrete SIM phase shifts. \label{fig:EE_vs_quant_bit}}
\end{figure}

To assess the impact of realistic imperfections, Fig. \ref{fig:EE_vs_quant_bit}
presents the EE of the proposed schemes for discrete SIM phase shifts.
The EE generally deteriorates as the number of quantization bits decreases,
especially for SIMs with a larger number of layers since the quantization
errors are amplified. As a rule of thumb, at least 3 bits per meta-element
are required for SIMs for a small $L$ (e.g., 2 or 4) to maintain
acceptable EE levels, while larger $L$ requires finer resolution.

\begin{figure}[t]
\centering{}\includegraphics[width=8.85cm]{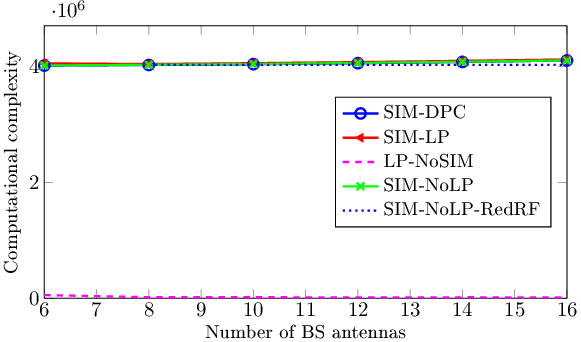}\caption{Computational complexity versus the number transmit antennas.\label{fig:Complexity}}
\end{figure}

\textcolor{black}{Finally, Fig. \ref{fig:Complexity} shows the per-iteration
computational complexity of the proposed optimization schemes versus
the number of transmit antennas. The LP-NoSIM scheme achieves the
lowest complexity compared to other schemes, as precoding without
SIM involves much fewer elements than SIM meta-elements. Due to the
same reason, the complexity of the benchmark schemes without precoding
is very similar to that of the proposed schemes, and it is not even
possible to observe any deference between DPC and LP. }
\begin{figure}[t]
\centering{}\textcolor{black}{\includegraphics[width=8.85cm]{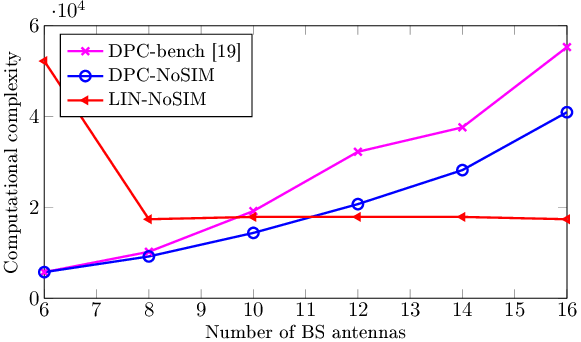}\caption{\textcolor{black}{Computational complexities of the proposed schemes
with DPC and LP that are without the SIM, and the scheme from \cite{xu2013energy}
versus the number transmit antennas.\label{fig:Complexity_wo_SIM}}}
}
\end{figure}

\textcolor{black}{To better highlight the difference between DPC and
LP, we compare the computational complexities of their appropriate
schemes without SIM (i.e., DPC-NoSIM and LP-NoSIM) in Fig. \ref{fig:Complexity_wo_SIM}.
As a benchmark, we consider the DPC-based scheme from \cite{xu2013energy},
also without SIM, referred to as }\textcolor{black}{\emph{DPC-bench.}}\textcolor{black}{{}
Its complexity is $O(I_{D}K(N_{t}^{2}N_{r}+N_{t}^{3}))$, where $I_{D}$
is the required number of loops. It can be observed that LP-NoSIM
has the highest complexity for $N_{t}=6$ and $N_{t}<KN_{r}$. As
$N_{t}$ further increases, the complexities of DPC-NoSIM and DPC-bench
become larger than that of LP-NoSIM. This is because the LP-NoSIM
complexity remains approximately constant for $N_{t}\ge KN_{r}$,
as indicated by (\ref{eq:Comp_lin_prec_X})). In addition, }\textcolor{black}{\emph{DPC-bench}}\textcolor{black}{{}
has a higher complexity than DPC-NoSIM, despite both achieving the
same energy efficiency.}

\section{Conclusion, Insights, and Future Work}

In this paper, we studied the EE maximization in a SIM-aided broadcast
system, considering both DPC and LP at the BS. For DPC, we exploited
the well-known BC-MAC duality to reformulate the optimization problem
and employed a \ac{SCA}-based technique to derive a tight lower bound
on the achievable sum-rate, which was then optimized using Dinkelbach\textquoteright s
method. A similar approach was used to optimize precoders in the case
of LP. The phase shifts of the SIM meta-elements were optimized using
a conventional projected gradient-based method due to its simplicity.
Also, we conducted a computation complexity analysis of the proposed
optimization algorithms and proved their convergence.

We carried out extensive numerical experiments and the simulation
results have lead to the following insights which are useful for designing
energy-efficient SIM-based systems:
\begin{itemize}
\item The proposed optimization algorithms significantly improve the EE
for SIM-aided MIMO broadcast systems.
\item DPC outperforms LP in terms of EE due to its superior interference
cancellation, but the EE improvement remains marginal, \emph{making
LP a practical alternative with lower complexity}.
\item SIM-based systems without digital precoding still suffer from higher
multi-user interference, leading to lower EE.
\item Increasing the number of meta-elements improves EE, but with diminishing
returns due to increased power consumption and the logarithmic nature
of the achievable sum-rate.
\item Optimal meta-element distribution across layers is crucial; simply
increasing the number of SIM layers can degrade performance due to
inter-layer signal propagation loss.
\item Discrete phase shift quantization negatively impacts EE, especially
for systems with more SIM layers.
\item In SIM-aided broadcast systems without digital precoding, optimal
energy efficient transmission strategies typically involve a subset
of active transmit antennas.
\end{itemize}
\textcolor{black}{For future work, the above insights immediately
motivate investigating the optimal distance between SIM layers to
achieve the best EE performance. Another interesting topic is studying
the performance of SIMs in a dynamic environment with user mobility,
which would involve adapting the proposed optimization algorithms.}

\appendix[Proof of Theorem \ref{thm:grad:phase:shift}]{}

Differentiating $R_{k}(\boldsymbol{\phi})$ in (\ref{eq:Rate_user})
w.r.t. $\mathbf{H}_{k}$ yields
\begin{equation}
\text{d}R_{\text{L},k}(\boldsymbol{\phi})=\text{d}\ln\left|\mathbf{F}_{1,k}\right|-\text{d}\ln\left|\mathbf{F}_{2,k}\right|
\end{equation}
where
\begin{align}
\!\!\!\text{d}\ln\left|\mathbf{F}_{1,k}\right| & =\tr\left(\hat{\mathbf{P}}_{s}\mathbf{H}_{k}^{H}\mathbf{F}_{1,k}^{-1}\text{d}\mathbf{H}_{k}+\mathbf{F}_{1,k}^{-1}\mathbf{H}_{k}\hat{\mathbf{P}}_{s}\text{d}\mathbf{H}_{k}^{H}\right)\\
\!\!\!\text{d}\ln\left|\mathbf{F}_{2,k}\right| & =\tr\left(\hat{\mathbf{P}}_{s}\mathbf{H}_{k}^{H}\mathbf{F}_{2,k}^{-1}\text{d}\mathbf{H}_{k}+\mathbf{F}_{2,k}^{-1}\mathbf{H}_{k}\hat{\mathbf{P}}_{k}\text{d}\mathbf{H}_{k}^{H}\right).
\end{align}
Substituting (\ref{eq:dHk}) into the previous expressions, we obtain
\begin{align}
\text{d}\ln\left|\mathbf{F}_{1,k}\right| & =\tr\left(\mathbf{G}_{1}^{H}\text{d}\boldsymbol{\mathbf{\Phi}}^{l}+\mathbf{G}_{1}\text{d}(\boldsymbol{\mathbf{\Phi}}^{l})^{H}\right)\\
\text{d}\ln\left|\mathbf{F}_{2,k}\right| & =\tr\left(\mathbf{G}_{2}^{H}\text{d}\boldsymbol{\mathbf{\Phi}}^{l}+\mathbf{G}_{2}\text{d}(\boldsymbol{\mathbf{\Phi}}^{l})^{H}\right)
\end{align}
where
\begin{align}
\mathbf{G}_{1}= & \boldsymbol{\Theta}^{l+1:L}\mathbf{G}_{k}^{H}\mathbf{F}_{1}^{-1}\mathbf{H}_{k}\hat{\mathbf{P}}_{s}\boldsymbol{\Theta}^{1:l-1}\mathbf{W}_{l}^{H}\\
\mathbf{G}_{2}= & \boldsymbol{\Theta}^{l+1:L}\mathbf{G}_{k}^{H}\mathbf{F}_{2}^{-1}\mathbf{H}_{k}\hat{\mathbf{P}}_{k}\boldsymbol{\Theta}^{1:l-1}\mathbf{W}_{l}^{H}.
\end{align}
After a few simple mathematical steps, we get
\begin{gather}
\nabla_{\boldsymbol{\phi}^{l}}R_{\text{L},k}(\boldsymbol{\phi})=\vect_{d}\left(\boldsymbol{\Theta}^{l+1:L}\mathbf{G}_{k}^{H}\mathbf{F}_{1,k}^{-1}\mathbf{H}_{k}\hat{\mathbf{P}}_{s}\boldsymbol{\Theta}^{1:l-1}\mathbf{W}_{l}^{H}\right)\nonumber \\
-\vect_{d}\left(\boldsymbol{\Theta}^{l+1:L}\mathbf{G}_{k}^{H}\mathbf{F}_{2,k}^{-1}\mathbf{H}_{k}\hat{\mathbf{P}}_{k}\boldsymbol{\Theta}^{1:l-1}\mathbf{W}_{l}^{H}\right).
\end{gather}
From this gradient expression for the achievable rate of user $k$,
we can easily obtain the appropriate gradients for all other users.
After summing all these gradient expressions, we obtain (\ref{eq:phi_grad}).
This completes the proof.

\bibliographystyle{IEEEtran}
\bibliography{IEEEabrv,IEEEexample,references}

\begin{thebibliography}{10}
\providecommand{\url}[1]{#1}
\csname url@samestyle\endcsname
\providecommand{\newblock}{\relax}
\providecommand{\bibinfo}[2]{#2}
\providecommand{\BIBentrySTDinterwordspacing}{\spaceskip=0pt\relax}
\providecommand{\BIBentryALTinterwordstretchfactor}{4}
\providecommand{\BIBentryALTinterwordspacing}{\spaceskip=\fontdimen2\font plus
\BIBentryALTinterwordstretchfactor\fontdimen3\font minus
  \fontdimen4\font\relax}
\providecommand{\BIBforeignlanguage}[2]{{%
\expandafter\ifx\csname l@#1\endcsname\relax
\typeout{** WARNING: IEEEtran.bst: No hyphenation pattern has been}%
\typeout{** loaded for the language `#1'. Using the pattern for}%
\typeout{** the default language instead.}%
\else
\language=\csname l@#1\endcsname
\fi
#2}}
\providecommand{\BIBdecl}{\relax}
\BIBdecl

\bibitem{recommendation2023framework}
``Framework and overall objectives of the future development of imt for 2030
  and beyond,'' \emph{International Telecommunication Union (ITU)
  Recommendation (ITU-R)}, 2023.

\bibitem{ericssonMobileData}
\BIBentryALTinterwordspacing
``{Mobile data traffic outlook -- Ericsson Mobility Report},'' 2024, [Accessed
  05-07-2024]. [Online]. Available:
  \url{https://www.ericsson.com/en/reports-and-papers/mobility-report/dataforecasts/mobile-traffic-forecast}
\BIBentrySTDinterwordspacing

\bibitem{di2020smart}
M.~Di~Renzo \emph{et~al.}, ``Smart radio environments empowered by
  reconfigurable intelligent surfaces: How it works, state of research, and the
  road ahead,'' \emph{IEEE Journal on Selected Areas in Communications},
  vol.~38, no.~11, pp. 2450--2525, 2020.

\bibitem{gong2023holographic}
T.~Gong \emph{et~al.}, ``Holographic {MIMO} communications: Theoretical
  foundations, enabling technologies, and future directions,'' \emph{IEEE
  Communications Surveys \& Tutorials}, vol.~26, no.~1, pp. 196--257, 2024.

\bibitem{zappone2022energy}
A.~Zappone \emph{et~al.}, ``Energy efficiency of holographic transceivers based
  on {RIS},'' in \emph{GLOBECOM 2022-2022 IEEE Global Communications
  Conference}.\hskip 1em plus 0.5em minus 0.4em\relax IEEE, 2022, pp.
  4613--4618.

\bibitem{shlezinger2021dynamic}
N.~Shlezinger \emph{et~al.}, ``{Dynamic metasurface antennas for 6G extreme
  massive MIMO communications},'' \emph{IEEE Wireless Communications}, vol.~28,
  no.~2, pp. 106--113, 2021.

\bibitem{you2022energy}
L.~You \emph{et~al.}, ``Energy efficiency maximization of massive {MIMO}
  communications with dynamic metasurface antennas,'' \emph{IEEE Transactions
  on Wireless Communications}, vol.~22, no.~1, pp. 393--407, 2022.

\bibitem{liu2022programmable}
C.~Liu \emph{et~al.}, ``A programmable diffractive deep neural network based on
  a digital-coding metasurface array,'' \emph{Nature Electronics}, vol.~5,
  no.~2, pp. 113--122, 2022.

\bibitem{an2024two}
J.~An \emph{et~al.}, ``Two-dimensional direction-of-arrival estimation using
  stacked intelligent metasurfaces,'' \emph{arXiv preprint arXiv:2402.08224},
  2024.

\bibitem{nadeem2023hybrid}
Q.-U.-A. Nadeem \emph{et~al.}, ``{Hybrid digital-wave domain channel estimator
  for stacked intelligent metasurface enabled multi-user MISO systems},''
  \emph{arXiv preprint arXiv:2309.16204}, 2023.

\bibitem{wang2024multi}
Z.~Wang \emph{et~al.}, ``Multi-user {ISAC} through stacked intelligent
  metasurfaces: New algorithms and experiments,'' \emph{arXiv preprint
  arXiv:2405.01104}, 2024.

\bibitem{hassan2024efficient}
N.~U. Hassan \emph{et~al.}, ``Efficient beamforming and radiation pattern
  control using stacked intelligent metasurfaces,'' \emph{IEEE Open Journal of
  the Communications Society}, vol.~5, pp. 599--611, 2024.

\bibitem{an2023stackedmulti}
J.~An \emph{et~al.}, ``Stacked intelligent metasurfaces for multiuser downlink
  beamforming in the wave domain,'' \emph{arXiv preprint arXiv:2309.02687},
  2023.

\bibitem{papazafeiropoulos2024achievableStatistical}
A.~Papazafeiropoulos \emph{et~al.}, ``Achievable rate optimization for large
  stacked intelligent metasurfaces based on statistical {CSI},'' \emph{IEEE
  Wireless Communications Letters}, 2024, {Early Access}.

\bibitem{liu2024drl}
H.~Liu \emph{et~al.}, ``{DRL-based orchestration of multi-user MISO systems
  with stacked intelligent metasurfaces},'' \emph{arXiv preprint
  arXiv:2402.09006}, 2024.

\bibitem{li2024stacked}
Q.~Li \emph{et~al.}, ``Stacked intelligent metasurfaces for holographic {MIMO}
  aided cell-free networks,'' \emph{IEEE Transactions on Communications}, 2024,
  {Early Access}.

\bibitem{an2023stackedholo}
J.~An \emph{et~al.}, ``{Stacked intelligent metasurfaces for efficient
  holographic MIMO communications in 6G},'' \emph{IEEE Journal on Selected
  Areas in Communications}, vol.~41, no.~8, pp. 2380--2396, 2023.

\bibitem{perovic2024mutual}
N.~S. Perovi{\'c} and L.-N. Tran, ``Mutual information optimization for
  {SIM}-based holographic {MIMO} systems,'' \emph{arXiv preprint
  arXiv:2403.18307}, 2024.

\bibitem{xu2013energy}
J.~Xu and L.~Qiu, ``Energy efficiency optimization for {MIMO} broadcast
  channels,'' \emph{IEEE Transactions on Wireless Communications}, vol.~12,
  no.~2, pp. 690--701, 2013.

\bibitem{lin2018all}
X.~Lin \emph{et~al.}, ``All-optical machine learning using diffractive deep
  neural networks,'' \emph{Science}, vol. 361, no. 6406, pp. 1004--1008, 2018,
  {Supplementary Material}.

\bibitem{vishwanath2003duality}
S.~Vishwanath \emph{et~al.}, ``Duality, achievable rates, and sum-rate capacity
  of gaussian {MIMO} broadcast channels,'' \emph{IEEE Transactions on
  Information Theory}, vol.~49, no.~10, pp. 2658--2668, 2003.

\bibitem{tam2016successive}
H.~H.~M. Tam \emph{et~al.}, ``Successive convex quadratic programming for
  quality-of-service management in full-duplex {MU-MIMO} multicell networks,''
  \emph{IEEE Transactions on Communications}, vol.~64, no.~6, pp. 2340--2353,
  2016.

\bibitem{perovic2022maximum}
N.~S. Perovi{\'c} \emph{et~al.}, ``On the maximum achievable sum-rate of the
  ris-aided mimo broadcast channel,'' \emph{IEEE Transactions on Signal
  Processing}, vol.~70, pp. 6316--6331, 2022.

\bibitem{zhao2023rethinking}
X.~Zhao \emph{et~al.}, ``{Rethinking WMMSE: Can its complexity scale linearly
  with the number of BS antennas?}'' \emph{IEEE Transactions on Signal
  Processing}, vol.~71, pp. 433--446, 2023.

\bibitem{perovic2021achievable}
N.~S. Perovi{\'c} \emph{et~al.}, ``{Achievable rate optimization for MIMO
  systems with reconfigurable intelligent surfaces},'' \emph{IEEE Transactions
  on Wireless Communications}, vol.~20, no.~6, pp. 3865--3882, 2021.

\bibitem{he2013coordinated}
S.~He \emph{et~al.}, ``Coordinated beamforming for energy efficient
  transmission in multicell multiuser systems,'' \emph{IEEE Transactions on
  Communications}, vol.~61, no.~12, pp. 4961--4971, 2013.

\bibitem{wang2024reconfigurable}
J.~Wang \emph{et~al.}, ``Reconfigurable intelligent surface: Power consumption
  modeling and practical measurement validation,'' \emph{IEEE Transactions on
  Communications}, 2024, {Early Access}.

\bibitem{huang2019reconfigurable}
C.~Huang \emph{et~al.}, ``Reconfigurable intelligent surfaces for energy
  efficiency in wireless communication,'' \emph{IEEE Transactions on Wireless
  Communications}, vol.~18, no.~8, pp. 4157--4170, 2019.

\bibitem{papazafeiropoulos2024achievable}
A.~Papazafeiropoulos \emph{et~al.}, ``{Achievable rate optimization for stacked
  intelligent metasurface-assisted holographic MIMO communications},''
  \emph{arXiv preprint arXiv:2402.16415}, 2024.

\end{thebibliography}

\end{document}